\documentclass{article}

\usepackage{amsmath,amssymb,graphicx,hyperref}

\usepackage{hyperref}
\usepackage{xcolor}
\usepackage{float}
\usepackage{graphicx}
\usepackage{titlesec}
\usepackage{placeins}
\usepackage{enumitem}
\usepackage{csquotes}
\usepackage{amsfonts}
\usepackage{amsmath}
\usepackage{amssymb}
\usepackage{enumitem}
\usepackage{booktabs}
\usepackage[numbers, compress, square, comma]{natbib}
\usepackage[range-phrase={\text{\ensuremath{-}}}]{siunitx}
\usepackage{subfig}
\usepackage{colortbl}
\usepackage{tikz}
\usepackage{tablefootnote}
\usepackage{algorithm}
\usepackage{algpseudocode}
\usepackage{makecell}
\usepackage[top=1.5in, bottom=1.5in, left=1.5in, right=1.5in]{geometry}
\usepackage[resetlabels]{multibib}
\newcites{AC}{References}
\newcites{AD}{References}
\newcites{AE}{References}
\usepackage{array}
\usepackage{multirow}

\usepackage{longtable}
\usetikzlibrary{shapes.geometric}

\usetikzlibrary{positioning}
\titleformat{\paragraph}
{\normalfont\normalsize\bfseries}{\theparagraph}{1em}{}
\titlespacing*{\paragraph}
{0pt}{3.25ex plus 1ex minus .2ex}{1.5ex plus .2ex}

\newcolumntype{L}[1]{>{\raggedright\let\newline\\\arraybackslash\hspace{0pt}}m{#1}}
\newcolumntype{C}[1]{>{\centering\let\newline\\\arraybackslash\hspace{0pt}}m{#1}}
\newcolumntype{R}[1]{>{\raggedleft\let\newline\\\arraybackslash\hspace{0pt}}m{#1}}
\usepackage[textsize=footnotesize]{todonotes}

\def\MELD{\texttt{MELD}}
\def\s90{S_{90}}

\newcommand\T{{\hspace{-.25pt}\intercal}}

\usepackage{booktabs}
\usepackage{longtable}
\usepackage{array}
\usepackage{multirow}
\usepackage{wrapfig}
\usepackage{float}
\usepackage{colortbl}
\usepackage{pdflscape}
\usepackage{tabu}
\usepackage{threeparttable}
\usepackage{threeparttablex}
\usepackage[normalem]{ulem}
\usepackage{makecell}
\usepackage{xcolor}

\title{A discrete event simulator for policy evaluation in deceased-donor liver allocation in Eurotransplant}
\author{H.C. de Ferrante \and M. de Rosner-Van Rosmalen \and B. Smeulders \and F.C.R. Spieksma \and S. Vogelaar}
\date{\today}

\begin{document}

\maketitle


\begin{abstract}
We present the ELAS simulator, a discrete event simulator built for the 
Eurotransplant (ET) Liver Allocation System (ELAS). Eurotransplant uses ELAS 
to allocate deceased-donor livers in eight European countries. The simulator
is publicly available; this provides transparency concerning the models used 
by Eurotransplant to evaluate liver allocation policies, and facilitates 
collaborations with policymakers, scientists, and other stakeholders in 
evaluating alternative liver allocation policies. This paper describes the
design and modules of the ELAS simulator. One of the included modules is the
obligation module, which is instrumental in ensuring that international 
cooperation in liver allocation benefits all ET member countries. 
\par 
By default, the ELAS simulator simulates liver allocation according to the 
actual Eurotransplant allocation rules. Stochastic processes, such as graft 
offer acceptance behavior and listing for a repeat transplantation, are 
approximated with statistical models that were calibrated using data from 
the Eurotransplant registry. We validate the ELAS simulator by comparing
simulated waiting list outcomes to historically observed waiting list outcomes
between 2016 and 2019.
\par 
The modular design of the ELAS simulator gives end users maximal control 
over the rules and assumptions under which Eurotransplant liver allocation 
is simulated, which makes the simulator useful for policy evaluation. We 
illustrate this with two clinically motivated case studies, for which we 
collaborated with hepatologists and transplantation surgeons from two liver
advisory committees affiliated with Eurotransplant.
\end{abstract}

\newpage
\normalsize

\section{Introduction}\label{introduction}

Eurotransplant (ET) is responsible for the allocation of deceased-donor
organs in eight European countries (Austria, Belgium, Croatia, Germany,
Hungary, Luxembourg, the Netherlands, and Slovenia). For liver
allocation, an important role is played by the Eurotransplant Liver and
Intestine Advisory Committee (ELIAC). This advisory committee monitors
the Eurotransplant liver allocation system (ELAS) and makes proposals
to the Eurotransplant Board on how to improve liver allocation rules
according to latest medical insights. ELIAC has brought forward several
issues with ELAS; for example, it has been reported that ELAS
overprioritizes candidates who receive exception points
\citep{umgelterDisparitiesEurotransplantLiver2017a}, and it has been reported
that Model for End-stage Liver Disease (MELD)
\citep{kamathModelPredictSurvival2001}, ET's basis for liver allocation, may
disadvantage female transplantation candidates
\citep{moylanDisparitiesLiverTransplantation2008, deFerranteSexDisparityLiver2024}.
Despite identification of such issues, the current liver allocation
system has not changed much since December 2006, when MELD scores became
the basis for liver allocation in Eurotransplant.

Reasons for the limited evolution of ELAS are questions on whether the
proposed modifications to allocation rules are adequate, and concerns
that proposed policy changes may have unintended consequences. These
doubts and concerns raise the need for a tool that can quantitatively
map how changes to allocation rules affect the mortality rates on the
liver waiting list and access to transplantation. There is a long
history of using operations research and discrete event simulation for
this purpose, with early work including the development of the UNOS
Liver Allocation Model (ULAM) for U.S. liver allocation \citep{Pritsker1995}
and use of computer simulation for design of the 1996 Eurotransplant
Kidney Allocation System
\citep{wujciakProposalImprovedCadaver1993a, demeesterNewEurotransplantKidney1998}.

This work has culminated in the development of a family of discrete
event simulators called the Simulation Allocation Models (SAMs)
\citep{ThompsonXSAM2004}, which are maintained by the Scientific Registry of
Transplant Recipients (SRTR) and tuned to allocation systems in the
United States. LSAM -- SRTR's tool for liver allocation
\citep{ThompsonXSAM2004} -- is used routinely by the scientific community and
policy makers to study alternative liver allocation rules. For example,
LSAM has been used to study the impact of expanding MELD with extra
biomarkers \citep{kimHyponatremiaMortalityPatients2008a, kimMELD3point0},
the impact of alternative geographic sharing rules
\citep{freemanImprovingLiverAllocation2004, axelrodEconomicImplicationsBroader2011, gentryAddressingGeographicDisparities2013, goelLiverSimulatedAllocation2018, akshatHeterogeneousDonorCircles2024},
and the impact of measures which improve access to transplantation for
specific patient groups such as pediatric and female patients
\citep{peritoImpactIncreasedAllocation2019, heimbachDelayedHepatocellularCarcinoma2015, bernardsAwardingAdditionalMELD2022}.
Other organ allocation organizations also routinely note the usage of
simulators for evaluation of new allocation policies, for example in
France \citep{jacquelinet2006changing, bayer2021removing}, the United
Kingdom \citep{watson2020overview}, the Netherlands \citep{deKlerk2021creating}
and India \citep{shoaibDiscreteeventSimulationModel2022}. More recently,
simulation-optimization has been used in the operations research
literature to design optimized allocation policies
\citep{bertsimasBalancingEfficiencyFairness2020b, papalexopoulosReshapingNationalOrgan2024, mankowskiRemovingGeographicBoundaries2023}.

Simulation can thus play a key role in moving Eurotransplant's liver
allocation system forward. However, existing simulation models are not
directly applicable to Eurotransplant. Such models typically simulate
allocation for a single patient population, whereas Eurotransplant needs
to balance the interests of the populations from its eight member
countries. A complicating factor in this is that Eurotransplant member
countries have different organ donation rates per million population
\citep{et_donor_pmp}, such that they face different challenges in liver
allocation. Moreover, existing models typically implement allocation
rules that are specific to the country for which the simulator was
designed. For instance, allocation rules of SAM software place much
emphasis on physical distances between the donor and transplantation
candidate, because geographical sharing in the United States is
primarily constrained by physical distances. In Eurotransplant, on the
other hand, geographical sharing is mostly impeded by country borders.
ELAS therefore gives substantial priority to candidates located in the
same country as the donor, and implements a mechanism that
ensures that livers transplanted with priority across country borders
are repaid to the exporting country. Existing simulation models do not
implement such mechanisms, and are thus a poor fit for Eurotransplant.

This has motivated us to develop a discrete event simulator tailored to
Eurotransplant. We refer to this simulator as the ``ELAS simulator''. Code for the
ELAS simulator is implemented in Python and made publicly available
together with synthetic data.\footnote{\url{http://github.com/hansdeferrante/Eurotransplant_ELAS_simulator}} By default, the ELAS simulator
simulates liver allocation according to Eurotransplant allocation rules.
The modular design of the simulator enables end users to use the
simulator for policy evaluation.

It is clear that the development of the ELAS simulation could not be
done, and was not done, in isolation. Multiple stakeholders were involved
in various phases of the project. In particular, members of the
Eurotransplant Liver and Intestine Advisory Committee (ELIAC),
who are hepatologists and transplant surgeons who represent
Eurotransplant's member countries, have given feedback on numerous
occasions on the conceptual design of the simulator; also at
Eurotransplant Annual Meetings, physicians and transplantation
coordinators have commented on various aspects of the simulator.

The paper is structured as follows. In Section \ref{sec:elassystem}, we give a
description of Eurotransplant's liver allocation system (ELAS). In
Section \ref{sec:elasdesign} we
discuss the design of the ELAS Simulator, and give an overview of the
general flow of the simulations. In Section
\ref{sec:elasmodules}, we
describe important modules of the ELAS simulator, each of which emulates
a key aspect of the liver allocation process. In Section
\ref{sec:elasvv},
we discuss verification and validation of the ELAS simulator. In Section
\ref{sec:elascasestudies}, we illustrate with two clinically
motivated case studies that the ELAS simulator is useful for policy
evaluation. We conclude and discuss in Section
\ref{sec:elasconclusion}.

\section{The Eurotransplant Liver Allocation System (ELAS)}\label{sec:elassystem}

We provide a simplified description of ELAS below; comprehensive descriptions
are available elsewhere
(see \citep{jochmansAdultLiverAllocation2017, ETLiverMan2025}). Fundamental to
ELAS is the laboratory MELD (lab-MELD) score, which quantifies a
candidate's 90-day waiting list mortality risk based on blood
measurements of serum bilirubin, serum creatinine, and the INR
\citep{kamathModelPredictSurvival2001}. The rationale for using lab-MELD
scores for liver allocation is that the lab-MELD score is a strong and
internationally accepted predictor of a candidate's 90-day mortality
risk. However, prioritization of candidates is not solely based on
lab-MELD scores, because certain patient groups would be underserved by
such an allocation. Specifically, ELAS also prioritizes candidates with:

\begin{itemize}
\item
  the High Urgency (HU) status that gives absolute international
  priority to candidates with acute liver failure,
\item
  the Approved Combined Organ (ACO) status that is given to
  candidates who require a combined transplantation of a liver with a
  heart, lung, pancreas, or intestine,
\item
  pediatric MELD (PED-MELD) scores that are assigned automatically to candidates
  of pediatric age,\footnote{Younger than 16 in Germany and younger than 18 in the
    other member countries until March 2025} and
\item
  Standard and non-standard exception (SE / NSE) MELD scores that can be awarded
  to patients who deserve priority for reasons other than a high
  short-term mortality risk and for whom the lab-MELD score does not
  represent clinical severity of the disease. These reasons include
  risk of disease irreversibility (for instance, hepatocellular
  carcinoma) and quality of life reasons (for instance, polycystic
  liver disease).
\end{itemize}

When a liver becomes available for allocation, Eurotransplant runs the liver
\emph{match algorithm} against a central database in which candidates for liver
transplantation are registered. This computer algorithm implements the
prioritization mechanisms of ELAS. Based on all waiting candidates, this
algorithm returns a list of candidates eligible to receive the liver graft.
This match list is ordered based on donor and
candidate characteristics, and the order determines the sequence in
which candidates are offered the liver graft by Eurotransplant. An
actual Eurotransplant liver match list for an adult blood group A donor
reported from the Netherlands is shown in Table \ref{tab:tab1}.

\begingroup
\setlength{\aboverulesep}{0.2ex}
\setlength{\belowrulesep}{0.3ex}

\begin{table}[h]
\caption{Example of a match list for an adult liver graft donated by a blood group A donor in the Netherlands. Under the restricted ABO blood group rules, candidates
  with blood group A and AB are eligible for this offer. At the moment of allocation the Netherlands has an obligation to return a blood group A liver graft to Croatia, resulting in Croatian candidates being ranked higher than Dutch candidates in elective tiers. This specific liver graft was declined by all Croatian and Dutch candidates and finally transplanted into a Belgian recipient.}
\label{tab:tab1}
\centering
\resizebox{1\ifdim\width>\linewidth\linewidth\else\width\fi}{!}{%

\begin{tabular}{lC{1.6cm}C{2.5cm}C{0.6cm}C{1cm}C{1cm}C{1cm}C{1cm}C{1cm}C{1.5cm}}
    \toprule
    \makecell{tier} & \makecell{offered to} & \makecell{candidate\\country} & \makecell{rank} & \makecell{match-\\MELD} & \makecell{lab-\\MELD} & \makecell{PED-\\MELD} & \makecell{(N)SE\\-MELD} & \makecell{cand.\\blood\\group} & \makecell{offer\\accepted?}\\
    \midrule
    HU & patient & Austria & 1 &  & 25 & 22 &  & AB & No\\
    \cmidrule{1-10}
    & \makecell{center\\(29 patients)} & Croatia & 2 &  &  &  &  &  & No\\
    \cmidrule{2-10}
    &  &  & 3 & 28 & 16 &  & 28 & A & No\\
    \cmidrule{4-10}
    &  &  & 4 & 22 & 22 &  &  & A & No\\
    \cmidrule{4-10}
    &  &  & 5 & 20 & 8 &  & 20 & A & No\\
    \cmidrule{4-10}
    &  &  & 6 & 20 & 20 &  &  & A & No\\
    \cmidrule{4-10}
    &  &  & 7 & 17 & 17 &  &  & A & No\\
    \cmidrule{4-10}
    &  &  & 8 & 17 & 17 &  &  & A & No\\
    \cmidrule{4-10}
    &  &  & 9 & 16 & 16 &  &  & A & No\\
    \cmidrule{4-10}
    &  &  & 10 & 15 & 15 &  &  & A & No\\
    \cmidrule{4-10}
    &  &  & 11 & 14 & 14 &  &  & A & No\\
    \cmidrule{4-10}
    &  &  & 12 & 14 & 14 &  &  & AB & No\\
    \cmidrule{4-10}
    &  &  & 13 & 14 & 14 &  &  & A & No\\
    \cmidrule{4-10}
    &  &  & 14 & 13 & 13 &  &  & A & No\\
    \cmidrule{4-10}
    &  &  & 15 & 13 & 13 &  &  & A & No\\
    \cmidrule{4-10}
    &  &  & 16 & 9 & 9 &  &  & A & No\\
    \cmidrule{4-10}
    &  &  & 17 & 9 & 9 &  &  & A & No\\
    \cmidrule{4-10}
    &  &  & 18 & 9 & 9 &  &  & A & No\\
    \cmidrule{4-10}
    &  &  & 19 & 8 & 8 &  &  & A & No\\
    \cmidrule{4-10}
    &  &  & 20 & 8 & 8 &  &  & A & No\\
    \cmidrule{4-10}
    &  & \multirow{-19}{*}{\raggedright\arraybackslash Netherlands} & 21 & 6 & 6 &  &  & A & No\\
    \cmidrule{3-10}
    \multirow{-21}{*}{\raggedright\arraybackslash elective} & \multirow{-20}{*}{\raggedright\arraybackslash patient} & Belgium & 22 & 35 & 35 &  &  & A & Yes\\
    \bottomrule
\end{tabular}
}
\end{table}

\endgroup
\FloatBarrier

At the highest level, the Eurotransplant match list order is based on
\emph{match tiers}. Candidates in higher tiers having priority over
candidates in lower tiers. The first ELAS tier consists of candidates
with the High Urgency (HU) status. The second tier consists of
candidates with the Approved Combined Organ (ACO) status. Candidates
with HU or ACO status are given absolute international priority in
Eurotransplant. A payback mechanism is used for grafts accepted internationally
in HU and ACO tiers. Specifically, international HU / ACO transplantations create an
\emph{obligation} for the receiving country to offer the next available liver
within the same blood group to the donor country until the obligation is
redeemed (see Section \ref{sec:elasobligations} for more information).
Candidates without an HU or ACO status are referred to as \emph{elective} candidates
in Eurotransplant. Such elective candidates are ranked in the remaining tiers.

Whether candidates appear on the match list and the rank at which they appear
is jointly determined by patient \emph{eligibility} criteria (blood
groups), \emph{ranking} criteria (MELD, pediatric status, donor/recipient blood
group combination), and \emph{filtering} criteria (patients can indicate that they
do not want to be considered for certain donors with allocation
profiles). These patient eligibility criteria, ranking criteria, and
filtering criteria are multi-factorial and differ by
Eurotransplant member country.

Common to all countries is that elective candidates listed in the same country
as the donor have priority over elective candidates in other countries. Other
factors that affect the ranking in ELAS include the combination of the donor and
candidate blood groups, whether the donor and/or candidate are pediatric,
whether the adult is low-weight (\textless55 kg), and the candidate's geographical
with respect to the donor.
The order of the match list in elective tiers is also affected by obligations.
For example, when the match list shown in Table \ref{tab:tab1} was created, the Netherlands had an
obligation to offer a blood group A liver back to Croatia. Consequently,
in the elective tier all Croatian candidates were ranked above Dutch
candidates. The most important factor for ranking candidates in elective
tiers is the \emph{match}-MELD score (see Table
\ref{tab:tab1}). This match-MELD score is the maximum of a candidate's lab-MELD
score and any received exception points (PED-MELD, SE-MELD, or NSE-MELD). We point
out that exceptional scores (SE-MELDs or NSE-MELDs) are valid only if allocation is
national, or if an offer was based on an obligation to pay back a liver.

Most offers in ELAS are \emph{recipient-driven}, meaning that
Eurotransplant offers the liver graft to a named candidate
\citep{jochmansAdultLiverAllocation2017}. Under specific circumstances,
centers may select any blood group compatible candidate from their
waiting list for transplantation. Eurotransplant refers to such offers as
\emph{center offers}. Offers to Croatia based on an obligation are an example of
center offers, which explains why a single offer to 29 Croatian patients appears
on the match list in Table \ref{tab:tab1}. The position of this center offer
on the match list is determined by the highest ranked elective candidate
listed in the center. We note that national regulations diverge on
when offers are center-driven. These rules are described in the Eurotransplant
liver manual (see \citep{ETLiverMan2025}).

The order of the match list determines the sequence in which Eurotransplant contacts
centers in \emph{standard allocation}. In cases where the loss of a transplantable graft
is anticipated, Eurotransplant can deviate from this allocation order (see
\citep{jochmansAdultLiverAllocation2017, ETLiverMan2025}). For example, Eurotransplant
is allowed to start an \emph{extended allocation} procedure 2 hours before the planned
explantation procedure (1 hour in Germany). Centers in the vicinity of the donor
center are then contacted, and have 30 minutes to propose up to two
candidates for transplantation. The proposed candidate with the highest rank
on the match list is then selected for transplantation. If extended allocation is unsuccessful, Eurotransplant can also offer
the graft to centers located further away from the donor center on a first-come-first-serve basis
with \emph{competitive rescue allocation}. In
total, 20\% to 25\% of liver grafts are transplanted through \emph{non-standard} allocation
mechanisms \citep{jochmansAdultLiverAllocation2017};
extended allocation accounts for the majority of this figure.

\section{The design of the ELAS simulator}\label{sec:elasdesign}

The ELAS simulator was designed to enable end users to assess the impact of changes
to liver allocation rules on waiting list mortality rates and access to
transplantation in Eurotransplant. We follow existing literature
\citep{shechterClinicallyBasedDiscreteEvent2005, Pritsker1995, ratcliffeSimulationModellingApproach2001, ThompsonXSAM2004}
in using Discrete Event Simulation (DES) for this purpose. With DES,
complex processes are analyzed by determining how system states are
affected by a sequence of discrete events. Within the ELAS simulator,
the system states are (i) the statuses of transplantation candidates
(whether they remain alive, their last known MELD score, their accrued
waiting time, and other information used in allocation), and (ii)
obligations to return a graft. The discrete events which affect these
system states are (i) patient events, which directly modify the
candidate's state (e.g.~updates to the MELD score), and (ii) liver donation
events, which generally lead to the transplantation
of a candidate and which may result in the creation or redemption of an
obligation to pay back a liver graft.

Apart from Eurotransplant allocation rules, which candidate is
transplanted at what moment is affected by several stochastic processes.
One such process is center acceptance behavior. Such behavior plays a
key role in liver allocation because transplantation centers frequently
decline offers of liver grafts that the centers deem unsuitable for
their candidate. Another stochastic process is that recipients of a
liver transplant can experience graft failure, and be enlisted for a
repeat transplantation. To accurately simulate outcomes of the
Eurotransplant liver allocation process, the ELAS simulator also has to
mimic these stochastic processes.

For discussion of the design of the ELAS simulator, we find it helpful
to distinguish between

\begin{enumerate}
\def\labelenumi{\arabic{enumi}.}
\item
  the \emph{organ allocation environment}, with which we refer to the
  overall setting in which allocation policies operate. This
  environment is deterministically defined by the simulation settings
  and input streams. Of key importance are the simulation input streams, which
  are the datasets that define the donors which become available for
  transplantation, the candidates who appear on the liver waiting
  list, and the donor and patient events that drive ELAS simulations.
  These input streams also specify the nested structure of agents in
  the ELAS simulations: donors and patients are nested in hospitals and
  transplantation centers, that are in turn nested in the
  Eurotransplant member countries. The organ allocation environment has
  to be specified in advance of any simulation. We discuss the
  requirements for input streams in Section
  \ref{sec:elasenvironment}. How simulations are
  initialized based on input streams is discussed in Section
  \ref{sec:elasinit}. How simulations proceed is
  illustrated in Section
  \ref{sec:elasoverview},
\item
  \emph{simulation modules}, that are implemented in Python code. These
  modules emulate key aspects of the liver allocation procedure. These
  processes include the generation of liver match lists according to
  ELAS' liver allocation rules, the simulation of graft offer acceptance
  behavior, ELAS' obligation system, and ELAS' exception point system.
  We discuss the implemented modules in Section
  \ref{sec:elasmodules}.
\end{enumerate}

With this overall design, the ELAS simulator is similar to LSAM which
has been used extensively to study organ allocation in the United
States. We also point out several differences. Most importantly, the
ELAS simulator is developed specifically for Eurotransplant, and takes
into account the multi-national setting; we include international
sharing rules and allow for allocation rules to differ per country.
Other major differences are that (i) in ELAS simulations centers can decline
liver offers for all their candidates, (ii) the ELAS simulator can
approximate outcomes of non-standard allocation while LSAM always places
livers through standard allocation, (iii) organ offer acceptance models specific
to pediatric candidates are included, (iv) splitting of liver grafts is
simulated, and (v) simulation of re-listing for repeat transplantation is
based on historical re-listing data.

\subsection{The organ allocation environment}\label{sec:elasenvironment}

To simulate liver allocation with the ELAS simulator, users must provide input
streams. These input streams consist of donor and patient information, which is
used by the ELAS simulator to construct the discrete events which drive ELAS
simulations. Users of the ELAS simulator are free to base these input streams on actual
registry data, reordered registry data, synthetic data, or combinations
thereof. Ideally, choices on which input stream to use are based on the
end user's research question of interest. For example, end-users
interested in evaluating the impact of small changes to allocation rules
may use historical data from the Eurotransplant registry for simulation,
while end-users interested in evaluating the effect of an expansion to
the donor pool may need to extend the historical donor pool with synthetic
donor data.

The input stream for donors must specify all administrative and medical information
that is required for liver allocation and for predicting liver offer acceptance and
post-transplant survival. Such information includes the
donor reporting date, the donor hospital and donor reporting center, the donor weight
and height, the donor blood group, and the donor death cause. The
information on the donor is static and reflects the state of the donor
on the date that they are reported to Eurotransplant.

Basic administrative and medical information necessary for allocation is also
required for the input streams for transplantation candidates. Such information includes the candidate's center of listing,
the candidate's disease group, the candidate's age and weight, and the
candidate's blood group. To identify candidates listing for a repeat
transplantation, a candidate identifier must be specified. Next to
this static candidate information, dynamic information is required on
the evolution of candidate's medical condition while they await liver
transplantation. For this, the ELAS simulator requires an input stream
of candidate status updates. These status updates specify when and how a
candidate's state changes while they wait on the Eurotransplant waiting
list. Examples of such status updates are new biomarker measurements,
reception of or upgrades to exceptional scores (NSE / SE / PED-MELD), changes to the candidate's waiting
list status (HU / ACO / NT status), and exit
statuses for the candidate (waiting list removals or waiting list
deaths). Because candidates regularly modify their \emph{allocation
profiles}, profiles were also implemented as status updates. With these
allocation profiles, centers can indicate that a candidate does not want
to receive offers from certain liver donors. For example, many
candidates specify that they do not want to be offered grafts from
donors above a certain age.

Discrete-event simulators for organ allocation require
complete knowledge on what would happen to a candidate if they would
remain on the waiting list until waiting list death or waiting list
removal. For candidates transplanted in reality, such information is
necessarily partly counterfactual; after all, transplantation prevents
us from observing what would have happened to the transplanted candidate
if they had remained on the waiting list. For other simulators, this
\emph{counterfactual status problem} was tackled by complementing the real
statuses of a transplanted candidate with statuses copied over from a
similar but not-yet-transplanted candidate
\citep{shechterClinicallyBasedDiscreteEvent2005, ThompsonXSAM2004}. In Appendix
\ref{app:imputation}, we propose a procedure to complete
status updates based on this idea.

We describe the full procedure for imputing counterfactual status updates in
Appendix \ref{app:imputation}. Briefly, this procedure first constructs a
\emph{risk set} for each transplanted
candidate. This is a set of candidates who
(i) remain on the waiting list, (ii) are similar to the candidate on a set of pre-specified characteristics, and (iii) face similar 90-day mortality risks
in the absence of transplantation. A complete status update trajectory can then be
constructed by (i) randomly sampling a candidate from the risk set, and
(ii) copying over the future status updates from the sampled candidate.
By running this procedure repeatedly, we construct for each candidate
multiple potential status update trajectories. This allows us to also
quantify the uncertainty in this counterfactual status imputation
procedure in simulations.

\subsection{Initialization of the ELAS simulator's system state}\label{sec:elasinit}

End users of the ELAS simulator must specify a simulation start and end
date, which jointly define the simulation time window. When simulation
starts, the system state is initialized by loading from input streams
all donors who were reported during the simulation window, and all candidates who
had an active waiting list status during the simulation window. For the
loaded candidates, all status updates are pre-processed until the
simulation start date. This ensures that the states of candidates at the
start of the simulation coincide with the status their actual status, as
specified in candidate input streams.

Candidates who (i) have in reality received a transplantation after the
simulation start date and (ii) were enlisted for repeat liver transplantation
before the simulation end date are potentially problematic, as
they could simultaneously await a primary and repeat transplantation in simulations.
To prevent these situations, the ELAS simulator by default ignores re-listings
of such candidates from the candidate input stream. Instead, patient re-listing
is simulated by the post-transplant module. This module simulates the post-transplant
survival and the potential re-listing of transplant recipients based on
candidate and donor characteristics (see Section \ref{sec:elasposttransplant}).

Initialization of the discrete-event simulation is finalized by
scheduling for each donor and transplantation candidate a single event
in the Future Event Set (FES). For donors, events are scheduled at the
donor reporting date as specified in the donor input stream. For each
candidate, a single patient event is scheduled at the time of the
candidate's first available status update after the simulation start
date. Subsequent updates are scheduled in the FES only after the
existing patient event has been handled.\footnote{This is necessitated by the
  fact that processing of a status update may result in automatic scheduling
  of a new update, see Section \ref{sec:elasexceptions}.}

\newpage

\subsection{Overview of the simulation}\label{sec:elasoverview}

\begin{figure}[h]

{\centering \includegraphics[width=0.9\linewidth]{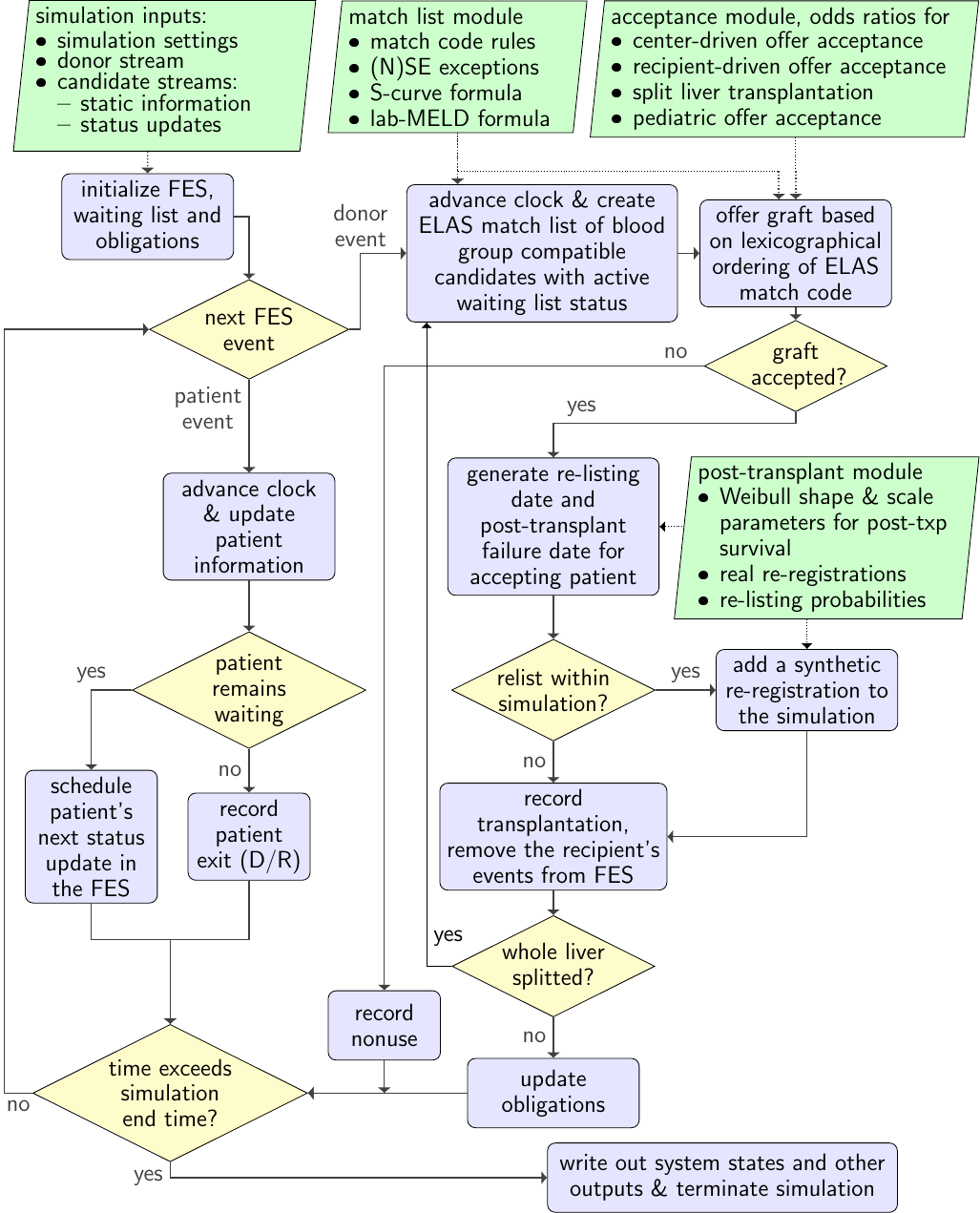} 

}

\caption{Event handling flowchart for the ELAS simulator. Inputs and parameters are represented using parallelograms. D, deceased on the waiting list; ELAS, Eurotransplant Liver Allocation System; FES, Future Event Set; MELD, Model for End-Stage Liver Disease; R, waiting list removal.}\label{fig:fig1}
\end{figure}

Figure \ref{fig:fig1} shows how patient and donor events from the
FES are processed in the ELAS simulator.
  The main categories of input parameters required for the simulation are shown in the green parallelograms in Figure \ref{fig:fig1}, and listed in detail in Supplementary Table \ref{tab:stab2}. 
In case of a patient event, the
corresponding candidate's status is updated according to their earliest
scheduled status update.

Figure \ref{fig:fig1} shows how patient and donor events from the
FES are processed in the ELAS simulator. In case of a patient event, the
corresponding candidate's status is updated according to their earliest
scheduled status update.

When a donor event is processed, a match list is created. To appear
on this match list, candidates must have an active waiting list status
(T, HU, or ACO) and must be compatible with the donor according to ELAS
blood group rules. The match list consists of patient-driven and/or
center offers. The order of the match list is deterministic and
follows the allocation rules specified in the organ allocation
environment. By default, actual ELAS allocation rules are followed,
which are described in Section \ref{sec:elasmatchlist}.

The ordered match list serves as an input to the graft offering module
(see Section \ref{sec:elasacceptance}). This module mimics the graft
offering process in Eurotransplant and returns the candidate who
accepts the liver offer, if any. We point out that it is not obvious
for which candidate the offered liver graft is accepted based on the
match list alone. One reason for this is that offers are often declined
by the transplant centers; in fact, over half of the transplanted liver
grafts in Eurotransplant were declined by 10 or more candidates before being accepted for transplantation. A second reason is that Eurotransplant may deviate
from the standard allocation procedure to prevent the loss of
transplantable organs (see Section \ref{sec:elassystem}). In simulating whether
a candidate accepts the offer, the module can simulate both standard and non-standard
allocation.

The ELAS simulator assumes that the candidate who accepts the liver
graft is transplanted, and removes scheduled events for this
transplanted candidate from the Future Event Set. If a liver was allocated
across country borders in HU or ACO tiers, the obligation system module creates
obligations (see Section \ref{sec:elasobligations}). For livers exported based
on an obligation, obligations are redeemed. The post-transplant module simulates
a time-until-liver-failure for each transplantation (see Section
\ref{sec:elasposttransplant}). In case the transplant recipient
is simulated to be enlisted for a repeat liver transplantation before
the candidate's death, the post-transplant module schedules a ``synthetic''
re-listing for the candidate.

The processing of patient and donor events continues until the
simulation end date is reached. At simulation termination, information
on transplantations is written to an output file. A list of discarded
grafts is also written to output files, as are the final states
of all candidates present in the simulation. These final states include
candidate exit statuses (waiting, waiting list death, transplanted, or
waiting list removal), as well as their last reported MELD scores. Such
information may be summarized to calculate the
statistics relevant for a specific research question. We chose to
write raw information to files rather than predefined summaries of
information to give end users maximum flexibility in reporting more
complicated statistics.

\FloatBarrier

\section{Modules of the ELAS simulator}\label{sec:elasmodules}

This section describes key modules of the ELAS simulator: the match list
module (Section \ref{sec:elasmatchlist}), the obligation system module (Section
\ref{sec:elasobligations}), the exception module (Section \ref{sec:elasexceptions}),
the graft offering module (Section \ref{sec:elasacceptance}), and the post-transplant module
(Section \ref{sec:elasposttransplant}).

\FloatBarrier

\subsection{The match list module}\label{sec:elasmatchlist}

When a liver graft becomes available for transplantation, the match list
module creates an ordered list of candidates who sequentially receive
offers from Eurotransplant until a candidate accepts the liver
graft. To appear on these match lists, candidates must (i) have an active
waiting list status (transplantable, HU, or ACO), and (ii) have a blood
group compatible with that of the donor according to ELAS
allocation rules. By default, the match list module collapses offers to
candidates eligible for center offers into a single center offer object,
whose rank is set equal to the rank of the highest ranked candidate who is
eligible for the center offer. The match lists returned by the match list module
are thereby ordered lists of center-driven and patient-driven offers,
coinciding with the structure of match lists Eurotransplant actually uses
for allocation (see the example match list in Table \ref{tab:tab1}).

The precise ordering returned by the match list module is based on liver
allocation rules, which have to be pre-specified as part of organ
allocation environment. The rules made available with the ELAS simulator
are the ELAS allocation rules used between 2016 and 2024. End-users of the
ELAS simulator can modify these allocation rules to study waiting list outcomes under
alternative allocation policies.

On a technical level, the ranking of ELAS match objects (patient- or
center offers) is based on \emph{match codes}. These match codes
consist of several components. The values of these components are determined
by Eurotransplant allocation rules, donor and patient characteristics, and the existence of obligations, and consist of:

\begin{enumerate}
\def\labelenumi{\arabic{enumi}.}
\item
  \textbf{Match tiers}, which are used to give absolute international
  priority to patients with HU status or ACO status,
\item
  \textbf{Match layers}, which differ per country and are used to give
  priority to certain patient groups (for example, to pediatric
  candidates, to blood group-identical candidates, to local
  candidates, and to candidates located in other countries based on
  obligations),
\item
  \textbf{Match obligation ranks}, which are used as a tiebreaker in case
  the donor country has obligations to return livers to multiple
  countries,
\item
  \textbf{Match MELDs}, which are used to rank candidates in the elective
  tiers,
\item
  \textbf{Match locality}, which is used to prioritize patients regionally
  in Germany in elective tiers,
\item
  \textbf{Waiting time}, which is the number of days with a HU or ACO status
  in the HU and ACO tiers. In elective tiers, this counts the number of
  consecutive days the candidate has had a match MELD at least as high
  as the current match MELD,
\item
  \textbf{The patient listing date}, which is used as a final tiebreaker.
\end{enumerate}

In ELAS, the order of the match list is based on the lexicographical ordering of these
components.

\subsection{The obligation system module}\label{sec:elasobligations}

To ensure a balanced exchange of liver grafts between Eurotransplant
member countries, ELAS includes an obligation system. Specifically,
grafts transplanted across country borders with international priority
(HU and ACO tiers) create an obligation for the recipient's country\footnote{
  In Austria, obligations are defined at the level of the
  transplantation center.}
to return an organ of the same blood group to the donor country.
An important feature of Eurotransplant's obligation system is that
obligations within the same blood group are automatically \emph{linked}. For
example, if the Netherlands has an obligation to return a blood group A
liver to Belgium, and an obligation is created for Belgium to return a
blood group A liver to Germany, the two existing obligations are
replaced by a \emph{linked} obligation for the Netherlands to return a blood
group A liver to Germany.

The obligation system module creates the obligations for grafts procured
internationally in HU / ACO tiers, and automatically merges linkable
obligations into a linked obligation. The module can also return
the outstanding obligations for a given blood group and country,
as well as the number of days these
obligations have existed. This information is required by the ELAS
simulator's match list module to determine match obligation ranks (see
Section \ref{sec:elasmatchlist}).

\subsection{The exception score module}\label{sec:elasexceptions}

Patient groups who are considered to be underserved by a purely
lab-MELD-based allocation can apply for standard (SE) or non-standard
(NSE) exceptions in ELAS. Pediatric patients automatically receive PED-MELD
scores based on their age. Candidates who receive such (N)SE and
PED-MELD scores are awarded predefined 90-day mortality equivalents,
which are specified in percentages. The exceptions, their eligibility criteria,
and their 90-day mortality equivalents vary by member country. For example,
in the Netherlands candidates with hepatocellular carcinoma are awarded a 10\%
mortality equivalent, whereas in Belgium a 15\% mortality equivalent is awarded.
For allocation, these mortality equivalents are translated to the MELD scale.
For example, a
10\% and 15\% mortality equivalent correspond to MELD scores of 20 and 22,
respectively \citep{ETLiverMan2025}. The awarded 90-day mortality equivalent
increases every 90 days\footnote{(N)SEs which have to be manually re-certified increase immediately upon re-certification.
  Re-certification is possible 14 days before an (N)SE expires, meaning that in practice, (N)SEs
  can increase every 76--90 days. In simulations, we assume that manually certificable
  (N)SEs are upgraded after 80 days. Certain (N)SEs and PED-MELD are automatically
  re-certified at expiry and increase every 90 days. Such (N)SEs are upgraded every
  90 days in simulations} for most (N)SEs and PED-MELDs, according to
exception- and country-specific increments. Some exceptions are implemented
as ``bonus SEs''. These bonus SEs add a fixed percentage mortality equivalent to the
lab-MELD's 90-day mortality equivalent.

The exception score module implements this system, and is initialized
based on an external file. This file has to specify which exceptions
exist, and relevant exception attributes (the initial mortality
equivalent awarded, the 90-day increment, the maximal equivalent
awarded, the maximum age after which the exception no longer increases,
and whether the SE is a regular SE or a bonus SE). For the default organ
allocation environment, all (N)SEs and PED-MELDs existing in 2023 were
implemented (see Appendix \ref{app:nses} for an overview). End-users may modify the attributes of these exceptions to
simulate Eurotransplant liver allocation under alternative (N)SE / PED-MELD rules. Simulation settings
also have to include parameters for the formula used to translate 90-day
mortality equivalents to the MELD scale.\footnote{This transformation is based on a Cox proportional hazards model which adjusts only for MELD and which uses a MELD score of 10 as the reference group. The curve used by Eurotransplant is given by: \[S_{90} = 0.98037 ^ {\exp\Big(0.17557(\texttt{UNOS-MELD} - 10)\Big)},\] where 0.98037 is the 90-day mortality equivalent for a MELD score of 10, and 0.17557 is the slope on MELD.}

By default, the ELAS simulator assumes that the candidate status input
stream also specifies when exceptions are upgraded or expire. In case no
future exception status is present in the candidate status queue for an
NSE / SE / PED-MELD, the ELAS simulator assumes that the candidate would
continue to re-certify their exception according to the
exception-specific re-certification schedule. This choice is motivated
by the fact that almost all candidates with exceptions re-certify them.

\subsection{The graft offering module}\label{sec:elasacceptance}

For a match list, the graft offering module returns the candidate who accepts the graft offer (if any), or indicates that all eligible candidates have rejected the offer. In case all eligible candidates have rejected the graft
offer, the ELAS simulator can either (a) force placement of the graft in
the candidate who was most likely to accept the graft offer, or (b)
record a discard.

The graft offering process is mimicked by (i) offering liver grafts to 
patients/centers in order of their ranking on the match list, and (ii) 
treating organ offer acceptance as a Bernoulli process in which graft offer 
acceptance probabilities are predicted based on a logistic model using
donor and patient characteristics (as in the SAM software, see \citep{SRTR2019}). 
Pseudocode for this graft offering process is included in Appendix
\ref{chap:supp}.
The graft offering module also includes a logistic model to predict whether a
liver graft is split by the transplantation center after acceptance.
Such split procedures allow centers to transplant two
candidates with one liver, typically one child and one adult.

There are several reasons why a candidate may not be offered a liver in
ELAS allocation despite being ranked high enough for an offer. These
reasons include that:

\begin{enumerate}
\def\labelenumi{\arabic{enumi}.}
\item
  Centers frequently decline the graft for all candidates who appear
  on the match list, even when making patient-driven offers (for
  example because of poor donor quality or capacity constraints). A
  patient may thus not receive a patient-driven offer in case the
  center has already rejected the graft.
\item
  Offers are not made to candidates whose allocation profiles exclude
  them from receiving the liver graft (for
  example because the donor is too old).
\item
  Center offers are not directly offered by Eurotransplant to individual candidates.
  Instead, the transplantation center chooses a candidate for transplantation.
\item
  Eurotransplant can deviate from the standard allocation procedure in
  case allocation time is limited. Offers
  to candidates not located in the vicinity of the graft are then
  bypassed.
\end{enumerate}

\noindent
These reasons motivated us to

\begin{enumerate}
\def\labelenumi{\arabic{enumi}.}
\item
  Implement a two-stage patient-driven offer acceptance procedure. In
  the first stage, a center-level logistic model is used to predict
  based on donor characteristics alone whether the center is willing
  to accept the graft. Provided that the center is willing to accept
  the graft, a patient-level logistic model is used to predict graft
  offer acceptance based on patient and donor characteristics.
\item
  Skip offers of liver grafts to elective candidates whose allocation
  profiles indicate that they do not want to be offered the liver graft.
\item
  Estimate logistic regressions separately for center- and
  patient-driven offers.
\item
  Approximate deviation from standard allocation. Specifically, a Cox
  proportional hazards model is used to simulate the number of offers
  made in regular allocation. Once this number of offers is reached,
  offers are only made locally (see Appendix
  \ref{app:devregacc} for details).
\end{enumerate}

Separate logistic models are used for four candidate groups:
(i) pediatric candidates
with HU / ACO statuses, (ii) elective pediatric patients, (iii) adult
candidates with HU / ACO statuses, and (iv) elective adult candidates.
This stratification is motivated by findings in the existing
literature that a single acceptance model poorly captures offer
acceptance behavior for specific patient groups. For example, Wood et al.
\citep{WoodPEDLSAM2021} note poor prediction for pediatric patients in LSAM.

To enable end users to change how graft offer acceptance decisions are
made, the odds ratios necessary for calculating graft offer acceptance
probabilities are kept in csv files external to the program. Default
odds ratios supplied with the ELAS simulator were estimated based on
offers of whole liver grafts offered between January 1, 2012 and December 31, 2019.
When estimating these odds ratios, we ignored offers automatically
rejected based on a candidate's allocation profile and offers accepted
in non-standard allocation. To account for correlations in organ
acceptance behavior, odds ratios were estimated with mixed effect models
with random effects for donor heterogeneity (as in
\citep{agarwalEquilibriumAllocationsAlternative2021}), patient heterogeneity,
and center heterogeneity.

\subsection{The post-transplant module}\label{sec:elasposttransplant}

Over 10\% of Eurotransplant liver waiting list registrations are
candidates who register for a repeat liver transplantation. To
accurately simulate such re-listings, a post-transplant module was
implemented for the ELAS simulator. At transplantation, this
post-transplant module first generates a time-to-failure \(T\), at which
the transplanted patient would die unless re-transplanted. This
time-to-failure is simulated based on a Weibull model, which accounts
for donor and patient characteristics. From this time-to-failure, we
find a time-to-re-listing \(R\) based on the empirical distribution of
re-listing times \(R\) relative to the time-to-failures \(T\).

In case the candidate is simulated to list for a repeat transplantation
within the simulation period, the post-transplant module constructs a
\emph{synthetic} re-listing. Such synthetic re-listings are generated by
combining the fixed patient characteristics of the transplant recipient
with the status updates from a candidate who was actually re-listed for
transplantation. This re-listing is chosen such that (a) candidates have
similar time-to-failure \(T\) and similar time-to-re-listing \(R\), and (b)
candidates are similar in terms of pre-determined characteristics (for
instance, both candidates are pediatric). One important matching
characteristic is whether the candidate re-lists within 14 days after
transplantation (\(R<14\)); this is motivated by the fact that many
candidates who experience graft failure within 14 days of
transplantation are eligible for an HU status per Eurotransplant
allocation rules. More details on how post-transplant survival and
re-listings are simulated is included in Appendix \ref{app:posttxp}.

The parameters required for simulation of post-transplant survival are
the shape and scale parameters for Weibull models, and the empirical
distribution of \(R\) (time-to-relisting) relative to \(T\)
(time-to-failure). The parameters supplied with ELAS simulator have been
estimated on Eurotransplant liver transplantations between 01-01-2012
and 31-12-2019. Since we expected post-transplant survival and
re-listing to be different for elective candidates and HU/ACO
candidates, we estimated Weibull parameters and curves separately for
these patient groups.

\section{Verification and validation of the ELAS simulator}\label{sec:elasvv}

We have described the design and the modules of the ELAS simulator in
Sections \ref{sec:elasdesign}
and \ref{sec:elasmodules}. In
this section, we describe our efforts to ensure that the ELAS simulator
adequately represents ELAS. For this, we distinguish between model
\emph{verification}, which refers to efforts taken to identify coding errors in the
software, and model \emph{validation}, which refers to efforts to assess whether
simulated outcomes closely approximate actual outcomes of Eurotransplant
liver allocation. For a general discussion of model verification and model
validation, we refer to Carson \citep{carsonVerificationValidationConsultants1989}.

\subsection{Verification of the ELAS simulator}\label{sec:elasverification}

To implement liver allocation rules in the ELAS simulator, we have used
the 2016 functional specifications for ELAS. To verify the correctness
of our implementation of ELAS allocation rules, we exported ELAS match
lists for 500 randomly selected donors who were reported between
January 1, 2016 and December 31, 2019. We constructed unit tests to verify that
the ELAS simulator correctly generated the match codes for these lists based
on the reported donor and candidate information. We also constructed unit tests to
ascertain that the ELAS simulator returns the candidates in the exact
same order as the exported match lists.

To ensure the correctness of the exception module, we exported (N)SE
exception score definitions directly from the Eurotransplant database.
For the obligation system module, we used functional specifications of
the ELAS obligation system to guide our implementation. These functional
specifications contain examples of how obligations have to be linked
including how to date obligations created from multiple linkable
obligations. We implemented these examples as unit tests for the ELAS
simulator.

\subsection{Validation of the ELAS simulator}\label{sec:elasvalidation}

Discrete-event simulators for organ allocation are typically validated
by comparing simulated statistics over a simulation window to real
statistics over the same time period
\citep{ThompsonXSAM2004, shechterClinicallyBasedDiscreteEvent2005}. Examples
of validated properties include the number of transplantations and the
number of waiting list deaths over the simulation window. Such
statistics are typically also presented stratified by patient
characteristics. We follow this practice to validate the ELAS simulator.

In comparing simulated statistics to observed statistics, we do not use
traditional hypothesis testing (or confidence intervals) because
\emph{``traditional hypothesis testing is not appropriate for measuring model
validity because the null hypothesis that the true system and model are identical is almost always false''}
\citep{shechterClinicallyBasedDiscreteEvent2005}. In fact, any difference
between actually observed outcomes and simulations can be made
statistically significant by increasing the number of simulation runs.
Instead, we give insight into the variability of ELAS simulator outputs
by reporting 95\% interquantile ranges (95\% IQRs) for simulation runs.
These interquantile ranges are obtained by simulating liver allocation
200 times, and reporting the 2.5th and 97.5th percentiles for simulated
outcomes of interest. When the real
summary statistic for an outcome of interest falls within the
interquantile range, we say that the ELAS simulator is
\emph{``well-calibrated''} for this outcome.

For input-output validation, we simulate Eurotransplant liver
allocation 200 times between January 1, 2016, and December 31, 2019.
The donor input stream consists of all donors reported in this period whose
livers were transplanted after allocation through ELAS. The candidate input stream consists of all patients
who had an active waiting list status (HU / T) in the simulation period.
We exclude candidates who were transplanted with a living
donor, because ELAS is only used to allocate deceased-donor livers, and
candidates whose country of listing changes (29 listings in total). For
simulation, we use actual donor arrival times and actual candidate
status histories, which were completed with the future status imputation
procedure described in Appendix \ref{app:imputation}. For each of the 200 simulation runs, a
different file with candidate status updates is used. These files differ
in the counterfactual status updates that were imputed for each
candidate.

Table \ref{tab:tab2} shows the validation results with
respect to waiting list outcomes. From Table \ref{tab:tab2}, it can
be seen that the ELAS simulator is well-calibrated for the number of
split liver transplantations, the number of listings, the number
of listings for a repeat transplantation, the number of waiting list
removals, and the number of waiting list deaths. A statistic for which
the simulator is not well-calibrated is the active waiting list size at
simulation termination, which is 4.5\% larger in simulations than
in reality. The ELAS simulator is well-calibrated to the number of
waiting list
deaths per country, except for in Germany where the number of waiting
list deaths is underestimated on average by 52 deaths (-8.1\%).
Inspecting waiting list deaths by lab-MELD shows that the ELAS simulator
underestimates the number of waiting list deaths in candidates with the
highest MELD scores (-8.1\% for MELD 31--40) and HU / ACO candidates
(-27.2\%), while the simulator overestimates the number of waiting list
deaths in candidates with low MELD scores (+11.8\% for MELD 6--10).

\begin{table}[h]
\caption{Validation of waiting list outcomes between January 1, 2016 and December 31, 2019. For simulations, the numbers shown are averages and 95\% interquantile ranges of outcomes over 200 simulations. Ranges are displayed in bold if the simulator is not well-calibrated, i.e., if the observed statistic does not fall within the 95\% IQR.}
\label{tab:tab2}
\centering
\resizebox{0.9\ifdim\width>\linewidth\linewidth\else\width\fi}{!}{%
\centering
\begin{tabular}{L{5.4cm} R{4.2cm} L{2.6cm}}
    \toprule
    \multirow[b]{1.5}{*}{category} & \multirow[b]{1.5}{*}{\makecell{simulated results\\(average and 95\%-IQR)}} & \multirow[b]{1.5}{*}{\makecell{actual data\\(2016-2019)}}\\
    & & \vspace{-0em} \\ 
    \midrule
    \addlinespace[0.3em]
    \multicolumn{3}{l}{\textbf{deceased-donor livers}}\\[.05cm]
    \hspace{1em}total transplantations & 6,415  [6,398-6,432] & 6,418 \\
    \hspace{1em}number of livers splitted & 173    [156-192] & 181 \\
    \hspace{1em}split transplantations & 346    [311-383] & 354 \\
    \addlinespace[0.3em]
    \multicolumn{3}{l}{\textbf{waiting list}}\\[.05cm]
    \hspace{1em}patient listings & 12,086 [12,034-12,141] & 12,110 \\
    \hspace{1em}relisting (synthetic) & 650    [598-705] & 652 \\
    \hspace{1em}final active waiting list & \textbf{1,528  [1,485-1,571]} & 1,462 \\
    \hspace{1em}removals (excl. recoveries) & 860  [831-1,888] & 857\\
    \hspace{1em}deaths & 1,636  [1,582-1,690] & 1,686 \\
    \addlinespace[0.3em]
    \multicolumn{3}{l}{\textbf{waiting list mortality by country}}\\[.05cm]
    \hspace{1em}Austria & 85     [72-100] & 82 \\
    \hspace{1em}Belgium & 169    [148-190] & 165 \\
    \hspace{1em}Croatia & 109    [97-121] & 98 \\
    \hspace{1em}Germany & \textbf{1,096  [1,051-1,140]} & 1,148 \\
    \hspace{1em}Hungary & 62     [52-72] & 72 \\
    \hspace{1em}Netherlands & 93     [79-107] & 101 \\
    \hspace{1em}Slovenia & 22     [17-29] & 20 \\
    \addlinespace[0.3em]
    \multicolumn{3}{l}{\textbf{waiting list mortality by lab-MELD}}\\[.05cm]
    \hspace{1em}lab-MELD 6-10 deaths & \textbf{123    [113-134]} & 110 \\
    \hspace{1em}lab-MELD 11-20 deaths & 498    [473-524] & 482 \\
    \hspace{1em}lab-MELD 21-30 deaths & 387    [361-413] & 379 \\
    \hspace{1em}lab-MELD 31-40 deaths & \textbf{505 [467-542]} & 546 \\
    \hspace{1em}HU / ACO deaths & \textbf{123    [103-142]} & 169 \\
    \bottomrule
\end{tabular}
}
\end{table}

Table \ref{tab:tab3} shows validation results relating
to transplantations, separately for standard and non-standard
allocation. The ELAS simulator is well-calibrated for most
summary statistics, including the number of placements through each
allocation mechanism, as well as the number of transplantations 
stratified by pediatric status and candidate sex. 
The simulator is not well-calibrated for the number
of transplantations within all Eurotransplant countries.

In total, 55 too many
transplantations ($+4.2\%$) are done in Belgium, while on average 35 (-5.6\%) and 15 (-3.0\%) too few transplantations are simulated in Austria
and Croatia, respectively. Restricting attention to standard allocation, we find that
too few grafts are accepted in Austria (-6.8\%), Croatia (-6.0\%) and Hungary (-7.4\%),
and too few grafts are accepted locally or regionally (-15\%).

Regarding transplantations categorized by type of exception, 
we find that the
ELAS simulator is well-calibrated for the number of transplantations in
non-exception candidates and in candidates with SE other than HCC.
The total number of transplantations with HCC is overestimated by 40 on average (+3.5\%), but is well-calibrated in standard allocation. The simulator
underestimates the number of transplantations in canddiates with NSEs,
both in standard (-17.5\%) and in total (-9.5\%).

In terms of the number of transplantations per MELD score, the ELAS
simulator appears to be well-calibrated for standard allocation, with
the number of transplantations underestimated for candidates with lab-MELD
6--10 (8.6\%) and overestimated for candidates with match-MELD 31--40
(+11.8\%). If we also include transplantations which result from non-standard
allocation, more differences emerge. In praticular, the ELAS simulator 
underestimates the number of transplantations in candidates with low 
MELD scores (MELD: 6–10) and overestimates the number of transplantations
in those with high MELD scores (MELD: 21–30 and 31–40).

\FloatBarrier

\begingroup
\setlength{\aboverulesep}{0.2ex}
\setlength{\belowrulesep}{0.3ex}

\begin{table}[b]
\caption{Validation of transplantations between January 1, 2016 and December 31, 2019. For simulations, the numbers shown are averages and 95\% interquantile ranges over 200 simulations. Ranges are displayed in bold if the observed statistic does not fall within the 95\% IQR.}
\label{tab:tab3}
\centering
\resizebox{1\ifdim\width>\linewidth\linewidth\else\width\fi}{!}{%
\hspace*{-.03\linewidth}
\centering
\setlength{\tabcolsep}{3pt}
\begin{tabular}{L{3.4cm} R{3.9cm} L{.9cm} R{3.9cm} L{.9cm}}
\toprule
\multicolumn{1}{c}{ } & \multicolumn{2}{c}{total transplantations} & \multicolumn{2}{c}{standard allocation only} \\
\cmidrule(l{3pt}r{3pt}){2-3} \cmidrule(l{3pt}r{3pt}){4-5}
\multirow[b]{1.5}{*}{category} & \multirow[b]{1.5}{*}{\makecell{simulated results\\(average and 95\%-IQR)}} & \multirow[b]{1.5}{*}{actual} & \multirow[b]{1.5}{*}{\makecell{simulated results\\(average and 95\%-IQR)}} & \multirow[b]{1.5}{*}{actual}\\
& & & & \vspace{-0em} \\

\midrule
\addlinespace[0.3em]
\multicolumn{5}{l}{\textbf{transplantations by allocation mechanism}}\\
\hspace{.7em}HU or ACO & 942    [905-977] & 931 & 935    [899-968] & 927\\
\hspace{.7em}obligation & \textbf{325    [302-354]} & 290 & \textbf{325    [302-354]} & 290\\
\hspace{.7em}MELD-based & 3850   [3636-4052] & 4005 & 3850   [3636-4052] & 4005\\
\hspace{.7em}extended or rescue & 1298   [1099-1503] & 1192 & \textemdash & \textemdash \\
\addlinespace[0.3em]
\multicolumn{5}{l}{\textbf{transplant recipients}}\\
\hspace{.7em}female & 2142   [2107-2186] & 2140 & 1764   [1690-1829] & 1825\\
\hspace{.7em}male & 4273   [4227-4310] & 4278 & 3346   [3196-3484] & 3397\\
\hspace{.7em}pediatric recipient & 418    [396-439] & 424 & 397    [372-417] & 405\\
\addlinespace[0.3em]
\multicolumn{5}{l}{\textbf{match geography}}\\
\hspace{.7em}local or regional & 2506   [2398-2600] & 2523 & \textbf{1577   [1496-1668]} & 1850\\
\hspace{.7em}national & 2664   [2543-2779] & 2647 & 2430   [2267-2581] & 2298\\
\hspace{.7em}international & 1245   [1183-1300] & 1248 & 1104   [1049-1155] & 1074\\
\hspace{.7em}Austria & \textbf{585    [561-605]} & 620 & \textbf{534    [508-563]} & 573\\
\hspace{.7em}Belgium & \textbf{1122   [1098-1142]} & 1067 & 997    [937-1040] & 983\\
\hspace{.7em}Croatia & \textbf{481    [466-493]} & 496 & \textbf{456    [437-470]} & 485\\
\hspace{.7em}Germany & 3178   [3148-3208] & 3166 & 2161   [1993-2323] & 2165\\
\hspace{.7em}Hungary & 298    [284-312] & 312 & \textbf{287    [270-301]} & 310\\
\hspace{.7em}Netherlands & 656    [626-681] & 656 & 583    [527-618] & 607\\
\hspace{.7em}Slovenia & 96     [88-103] & 101 & 92     [83-100] & 99\\
\addlinespace[0.3em]
\multicolumn{5}{l}{\textbf{transplantations by patient type (adult non-HU/ACO only)}}\\
\hspace{.7em}lab-MELD only & 3250   [3195-3299] & 3289 & 2453   [2327-2567] & 2482\\
\hspace{.7em}HCC & \textbf{1190   [1158-1221]} & 1150 & 903    [848-962] & 917\\
\hspace{.7em}NSE & \textbf{259    [239-275]} & 286 & \textbf{205    [179-224]} & 248\\
\hspace{.7em}other SE & 558    [533-579] & 547 & 421    [388-453] & 453\\
\addlinespace[0.3em]
\multicolumn{5}{l}{\textbf{transplantations by match-MELD (adult non-HU/ACO only)}}\\
\hspace{.7em}match-MELD 6--10 & \textbf{434    [400-476]} & 493 & 337    [304-368] & 360\\
\hspace{.7em}match-MELD 11--20 & 1388   [1294-1487] & 1466 & 970    [913-1021] & 982\\
\hspace{.7em}match-MELD 21--30 & \textbf{2482   [2386-2574]} & 2285 & 1792   [1633-1933] & 1769\\
\hspace{.7em}match-MELD 31--40 & \textbf{952    [899-1000]} & 1025 & \textbf{884    [826-935]} & 989\\
\hspace{.7em}unknown & -- & 3 &  & \\
\addlinespace[0.3em]
\multicolumn{5}{l}{\textbf{transplantations by lab-MELD (adult non-HU/ACO only)}}\\
\hspace{.7em}lab-MELD 6--10 & \textbf{1292   [1254-1337]} & 1373 & \textbf{974    [910-1028]} & 1066\\
\hspace{.7em}lab-MELD 11--20 & 2143   [2058-2226] & 2181 & 1535   [1458-1605] & 1553\\
\hspace{.7em}lab-MELD 21--30 & \textbf{1044   [993-1093]} & 942 & 752    [675-816] & 739\\
\hspace{.7em}lab-MELD 31--40 & 778    [731-828] & 769 & 721    [669-780] & 738\\
\hspace{.7em}unknown & -- & 7 & 0 & 4\\
\bottomrule
\end{tabular}
}
\end{table}

\endgroup

\FloatBarrier

\subsubsection{Validation of post-transplant outcomes}\label{sec:elasvalpost}

\begin{figure}[h]

{\centering \includegraphics[width=0.9\linewidth]{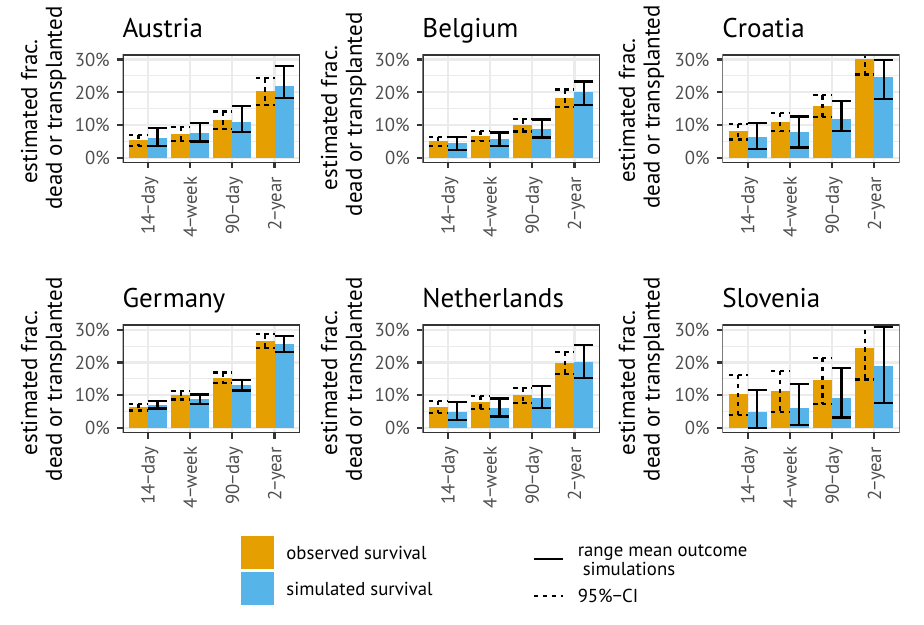} 

}

\caption{Estimated event probabilities for transplant recipients, per country for different time horizons. An event was defined as re-transplantation or waiting list death, whichever occurred first. Probabilities were estimated from real data and simulated data with the Kaplan-Meier estimator, using time since transplantation as the timescale.}\label{fig:fig2}
\end{figure}

Figure
\ref{fig:fig2} compares the simulated
post-transplant event rates with real post-transplant event rates,
estimated per country at several time horizons. For both, \(t\)-day
post-transplant survival probabilities are estimated with Kaplan-Meier.
Simulated event rates are close to real event rates in all
Eurotransplant regions. Only in Croatia and Slovenia, simulated
post-transplant event rates appear to be slightly biased downwards.
Supplementary Figure \ref{fig:sfig1} shows that estimated re-listing
probabilities are also comparable to observed re-listing probabilities.

\FloatBarrier

\subsubsection{Discussion of validation results}\label{sec:valelasdiscussion}

The ELAS simulator appears to be well-calibrated for most waiting list
and transplantation patterns within Eurotransplant. An important
exception is the total number of transplantations and waiting list exits 
per lab- and match-MELD score. For example, the simulator overestimates
the total number of transplantations among candidates with high urgency 
(MELD 31--40 or HU/ACO status), while the simulator underestimates the number of waiting list
deaths in this group (see Tables \ref{tab:tab2} and \ref{tab:tab3}).
Table \ref{tab:tab3} also shows that such miscalibration is to a much lesser degree present
in standard allocation, in which the number of transplantations is only
slightly overestimated for the candidates with the highest match-MELD scores (31--40)
and those with the lowest lab-MELD scores (6--10).

A potential explanation for this miscalibration is that the simulator 
assumes that candidates who accept a graft offer are always transplanted. 
In practice, however, transplantation centers may cancel transplantation
procedures after an initial acceptance, for example because of logistical
issues, unfavorably histopathology findings, or instability of the
transplantation candidate. In such cases, Eurotransplant can re-offer 
the graft to candidates who are located in the
vicinity of the graft via extended or rescue allocation. This generally results in
placement of livers in candidates with lower urgency, particularly in case of
rescue allocation because that operates on a first-come-first-serve basis.
We note that such rescue allocation was not implemented for the ELAS simulator.

To assess whether competitive rescue allocation could partially explain the
observed miscalibration, we inspected the placements of the 241 liver grafts that 
were initially accepted by one candidate, declined before transplantation, and
then re-offered to other candidates. For these 241 livers, 
Table \ref{tab:tabcomprescue} shows the MELD scores of the initially 
accepting candidate and final transplantation recipient. In general,
re-offering via rescue allocation indeed leads to lower MELD scores 
at transplantation. This indeed partially explains the observed miscalibration 
in low-MELD groups in Tables \ref{tab:tab2} and \ref{tab:tab3}. For example, in simulations there are on average 60 too few transplantations in 
candidates with match-MELD 6--10 (\ref{tab:tab2}), and re-allocation 
via rescue allocation leads to approximately 40 extra such transplantations.

\begin{table}[h]
\caption{Priority scores of transplant candidates in cases where a liver was initially accepted but later declined, and then re-allocated to a different recipient through competitive rescue liver allocation between January 1, 2016, and December 31, 2019. The table shows the number of candidates in each priority score category. For MELD scores, separate counts are included for the match-MELD score and lab-MELD score.}
\label{tab:tabcomprescue}
\centering
\resizebox{1\ifdim\width>\linewidth\linewidth\else\width\fi}{!}{%
\centering
\begin{tabular}{lcccc}
    \toprule
    score & \multicolumn{2}{c}{initially accepting candidate} & \multicolumn{2}{c}{final liver recipient} \\
    \midrule
    \addlinespace[0.3em]
    \multicolumn{5}{l}{\textbf{urgency category}}\\
    \hspace{.7em}HU/ACO         & \multicolumn{2}{c}{23} & \multicolumn{2}{c}{0}     \\
    \addlinespace[0.3em]
    \textbf{MELD} & match-MELD & lab-MELD & match-MELD & lab-MELD \\
    \hspace{.7em}31--40    & 60 & 38 & 1   & 0   \\
    \hspace{.7em}21--30    & 107 & 56 & 53  & 31  \\
    \hspace{.7em}11--20    & 45 & 85 & 140 & 157 \\
    \hspace{.7em}6--10     & 6  & 39 & 47  & 53  \\
    \bottomrule
\end{tabular}
}
\end{table}

A second outcome on which the simulator is not well-calibrated is the
number of transplantations per country. For example, in Belgium there
are on average 55 (+4\%) more transplantations in total in the simulations 
than in reality, while no miscalibration is observed in 
standard allocation. This miscalibration also seems to be explained by
lack of competitive rescue allocation, through which approximately 10--15
grafts are transferred per year from Belgium to Germany. The fact that
too many grafts are transplanted in Belgium may also explain why the
number of transplantations in candidates with HCC is overestimated in
total, as Belgium has the highest share of candidates listed with HCC.
This seems to be supported by the fact that no 
miscalibration is observed when focusing on standard allocation only 
(see Table \ref{tab:tab2}).

We thus have two potential explanations for miscalibration of the ELAS
simulator, which are (i) the simulator does not allow planned
transplantation procedures to be cancelled after an initial acceptance,
and (ii) the simulator does not implement competitive rescue allocation.
We have chosen not to implement these two mechanisms in the ELAS
simulator because (i) they are not directly relevant to standard ELAS
allocation, which is the focus of policy discussions in ELIAC, (ii) they
occur relatively rarely (less than 4\% of transplantations), and (iii)
the decision to proceed with rescue allocation is made on a case-by-case
basis for which limited structured data available, making it complex to
model correctly. All in all, we believe that the
simulated outcomes are sufficiently close to actual outcomes to make the
ELAS simulator useful for policy evaluation. We illustrate this with two
case studies in Section \ref{sec:elascasestudies}.

\section{Case studies: the impact of modifying ELAS allocation rules}\label{sec:elascasestudies}

We illustrate how the ELAS simulator can be used to quantify the impact of
policy changes with two case studies. For the first case study Section
\ref{sec:elasbeliaccasestudy}), we
collaborated with representatives from the Belgian Liver and Intestine
Advisory committee (BeLIAC) to study the impact of changes to the
Belgian exception score system. In the second case study (Section \ref{sec:elasremeldcasestudy}), we study the
impact of basing Eurotransplant liver allocation on ReMELD-Na scores
instead of UNOS-MELD scores, which was a topic on the agenda of ELIAC in
2023.

To evaluate the alternative policies, we simulate Eurotransplant liver
allocation 50 times between January 1, 2016, and December 31, 2019, and compare
simulated outcomes under modified ELAS rules to simulated outcomes under current rules.
We test whether modified policies lead to
significantly different outcomes with traditional hypothesis testing. To
increase the power of these tests, we use common random number
generators \citep{lawSimulationModelingAnalysis2015} to eliminate the variance attributable to factors which we
assume to be independent from the allocation policy. Specifically, we use common random
numbers to synchronize the splitting of liver grafts, the graft offer
acceptance behavior of candidates, and the triggering of rescue
allocation across policies for each of the 50 iterations. By
synchronizing these processes across policies the outcomes under alternative
policies may be compared to the current policy with pairwise t-tests.

We note that these simulations assume the statistical models used to approximate 
stochastic processes -- such as graft-offer acceptance behavior and post-transplant
survival -- remain unchanged across policy scenarios. In other words, we do not model 
potential behavioral adaptations by transplantation centers or changes in 
clinical outcomes that might arise in response to a policy shift. For the BeLIAC
case study, this means that the offer acceptance behavior of centers is not affected
by the fact that the exception scores are slowed down or capped. For the ReMELD-Na
case study, this implies that candidates with a specific lab-MELD score make
the same decisions regardless of whether this is a ReMELD-Na score or a UNOS-MELD
score. 

\subsection{Case study 1: the exception score system in Belgium}\label{sec:elasbeliaccasestudy}

In Belgium, nearly half of the liver transplantation candidates receive exception
points, which makes Belgium the member country with most awarded (N)SEs.
Most of the (N)SEs increase with every 90 days of waiting time, which
increases the match-MELD score candidates need in Belgium to receive
graft offers. This has led to concerns that candidates without
exception points are crowded out of transplantation.

To address this, the BeLIAC has considered imposing a cap on (N)SE-MELDs
of 30, which corresponds to maximizing the awardable 90-day mortality
equivalent with a 50\% mortality equivalent. In joint discussions, the
BeLIAC also expressed an interest in capping (N)SE-MELDs at 25, as well as
alternative policy options. These alternatives were slowing down (N)SE-MELDs by
reducing the 90-day increments (referred to as ``\emph{slower}'' policies), and lowering
the initial mortality equivalents awarded for (N)SEs (referred to as
``\emph{lowered}'' policies). Table \ref{tab:tab4} provides an overview
of the policy options discussed within BeLIAC.

\begin{table}[h]
\caption{Policy options for Belgian (N)SE system. Modifications are applied only to all Belgian SEs and the NSE, not the PED-MELD score (which is valid internationally). A dash (“-”) indicates that the policy did not change an (N)SE attribute.}
\label{tab:tab4}
\centering
\resizebox{1\ifdim\width>\linewidth\linewidth\else\width\fi}{!}{%
\begin{tabular}{C{.18\linewidth} C{.35\linewidth} C{.3\linewidth} C{.28\linewidth}}
    
        \toprule
        \rowcolor[HTML]{FFFFFF} 
        policy option        & initial   equivalent                                                                  & 90-day   increment                            & maximum mortality  equivalent \\ \midrule
        \rowcolor[HTML]{EFEFEF} 
        current & 10\% (MELD 20) for most (N)SEs, 15\%  (MELD 22) for HCC                                                    & 10\%     (2-4 MELD points)                  & 100\%                \\
        \rowcolor[HTML]{FFFFFF} 
        capped (25)  & -  & - & 25\%$^a$ (MELD 25)     \\
        \addlinespace[0.3em]
        \rowcolor[HTML]{FFFFFF} 
        capped   (30) & - & - & 50\%$^a$   (MELD 30)   \\[.1em]
        \rowcolor[HTML]{EFEFEF} 
        slower        & - & 5\%       (1-2 MELD points)                 & -                    \\[0.3em]
        \rowcolor[HTML]{EFEFEF} 
        slowest       & - & 2.5\%   (1 MELD point)                      & -                    \\
        \rowcolor[HTML]{FFFFFF} 
        lowered       & lowered to 8\% (MELD 18) for existing (N)SEs with initial equivalents \textless{}20\% & -                                           & -                    \\ \bottomrule
    \end{tabular}
}
\parbox{\textwidth}{\footnotesize \smallskip $^a$Set to the initial equivalent if the initial equivalent exceeds the proposed cap. For example, candidates with biliary atresia maintain a 60\% mortality equivalent.}
\end{table}

\begin{figure}[h]

{\centering \includegraphics[width=1\linewidth]{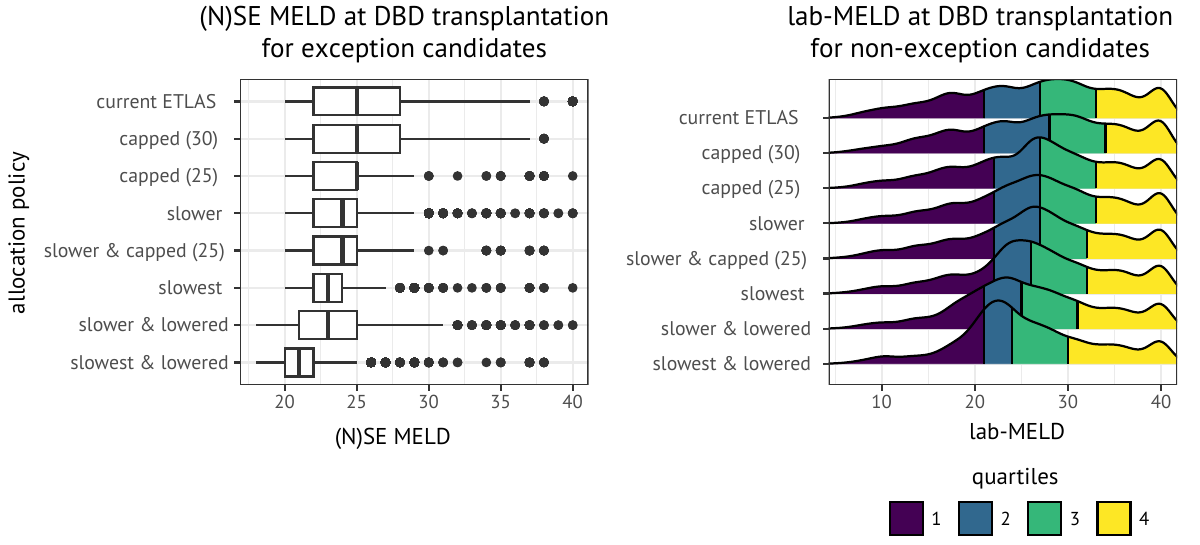} 

}

\caption{Distribution of (N)SE MELD scores for exception candidates (left) and laboratory MELD scores for non-exception candidates (right) at transplantation in ELAS simulations. Only transplantations based on recipient-driven offers are shown. Boxplots and distributions were calculated over 50 simulations.}\label{fig:fig3}
\end{figure}

\FloatBarrier

Results of the simulated impact of these policies are summarized in Figure
\ref{fig:fig3} and Table \ref{tab:tab5}, which both focus on livers
from DBD donors because DCD donors are center offers in Belgium. Figure
\ref{fig:fig3} visualizes the distributions of the simulated
(N)SE-MELD scores at DBD transplantation for exception candidates (left),
and laboratory MELD scores at DBD transplantation for non-exception
candidates (right). The left side shows that capping (N)SE-MELDs at 30
barely affects the distribution of (N)SE-MELDs at transplantation, while the other
policies reduce the median (N)SE-MELD at transplantation by up to 4 MELD points.
The right side shows that lab-MELD scores at DBD transplantation for non-exception
candidates also become lower when (N)SEs are capped. In fact,
with \emph{slower} and \emph{lowered} policies, the median lab-MELD at DBD transplantation
becomes up to 3 MELD points lower than under the current allocation rules. These alternative policies thus result in increased access to
transplantation for candidates with lab-MELD scores between 20 and 23,
which may be desirable because these candidates face 10--15\% 90-day
mortality risks.

Table \ref{tab:tab5} shows summary statistics for waiting list outcomes,
separately for exception and non-exception candidates. Comparing the third column (actual 2016--2019 statistics) to
the fourth column (averages and 95\% IQRs over the 50 simulations) shows that
the ELAS simulator is well-calibrated for the number of
transplantations, waiting list deaths, and waiting list exits. This
reassures us that the ELAS simulator reasonably describes these
allocation patterns. Of policy interest are the remaining columns of
Table \ref{tab:tab5}, which show for the (N)SE
policy alternatives the average outcome over the 50 simulations. Comparing
these outcomes to the outcomes simulated under current (N)SE rules shows
that almost all policies significantly change the number of
transplantations, the number of waiting list deaths, and the number of
waiting list removals.

\begingroup
\sisetup{range-phrase = {\,\textemdash\,}}

\begin{table}[h]
\caption{Waiting list exits for elective candidates in Belgium between January 1, 2016, and January 1, 2020. The numbers displayed are the average number of exits over 50 simulations. Scenarios modifying the allocation rules were compared to simulations under the current ELAS rules with pairwise t-tests.}
\label{tab:tab5}
\centering
\hspace*{-.05\linewidth}
\resizebox{1.1\ifdim\width>\linewidth\linewidth\else\width\fi}{!}{%
\begin{tabular}{C{.18\linewidth}C{.08\linewidth}C{.07\linewidth}C{.18\linewidth}C{.08\linewidth}C{.08\linewidth}C{.08\linewidth}C{.13\linewidth}C{.08\linewidth}C{.08\linewidth}C{.08\linewidth}}
    \toprule
    exit reason & \makecell{patient\\type} & \makecell{current\\ (real)} & \makecell{current\\ (sim)} & \makecell{capped\\(30)} & \makecell{capped\\(25)} & slower & \makecell{slower \&\\ capped (25)} & slowest & \makecell{slower \&\\ lowered} & \makecell{slowest \&\\ lowered}\\
\midrule
 & (N)SE & 438 & \phantom{}\num{450} [\numrange{428}{470}] & 445 & 448 & 443*** & 445* & 440*** & 438*** & 433***\\

\multirow{-2}{*}{\raggedright\arraybackslash transplanted} & None & 387 & \phantom{}\num{405} [\numrange{379}{423}] & 406* & 408** & 412*** & 409*** & 413*** & 420*** & 425***\\
\cmidrule{1-11}
 & (N)SE & 27 & \phantom{0}\num{27} [\numrange{19}{40}] & 28 & 27 & 28 & 29 & 29* & 29 & 29*\\

\multirow{-2}{*}{\raggedright\arraybackslash \makecell{waiting list\\death}} & None & 119 & \phantom{}\num{126} [\numrange{108}{141}] & 126 & 122** & 122* & 122** & 117*** & 116*** & 114***\\
\cmidrule{1-11}
 & (N)SE & 20 & \phantom{0}\num{14} [\numrange{8}{17}] & 14 & 14* & 14** & 14* & 15*** & 16*** & 17***\\

\multirow{-2}{*}{\raggedright\arraybackslash unfit} & None & 23 & \phantom{0}\num{21} [\numrange{16}{29}] & 21 & 20 & 21 & 20 & 20 & 20 & 20*\\
\cmidrule{1-11}
 & (N)SE & 8 & \phantom{00}\num{8} [\numrange{5}{16}] & 9 & 10** & 10** & 11*** & 11*** & 11*** & 13***\\

\multirow{-2}{*}{\raggedright\arraybackslash \makecell{removed\\HCC or cancer}} & None & 6 & \phantom{00}\num{6} [\numrange{2}{14}] & 6 & 6 & 6 & 6 & 6 & 6 & 6\\
\bottomrule
\end{tabular}
}
\parbox{\textwidth}{\footnotesize \smallskip $^{*}$p < 0.05; $^{**}$p < 0.01; $^{***}$p < 0.001}
\end{table}

\endgroup
\FloatBarrier

The greatest effects are sorted by the combined \emph{slowest} and \emph{lowered} policy
(final column), with which on average 20 (+5\%) extra candidates without exception
points are transplanted. According to the simulation results, this policy could
have avoided 12 waiting list deaths in non-exception candidates (-10\%), which corresponds to 3 waiting list
deaths per year. The cost of this policy change is that we see a slight increase
in the number of waiting list removals for exception patients, with on average
1--2 extra exception patients per year removed because they became unfit for
transplantation, or because they had HCC / cancer. This information can be used
to inform BeLIAC discussions on how the exception point system should be revised.

\subsection{Case study 2: basing Eurotransplant liver allocation on ReMELD-Na}\label{sec:elasremeldcasestudy}

Hyponatremic candidates (i.e., candidates with low serum sodium levels) face systematically
higher waiting list mortality rates than is suggested by
their MELD score \citep{kimHyponatremiaMortalityPatients2008a}. To adequately
prioritize these patients, MELD-Na was introduced in 2016 for liver allocation
in the United States. In 2020, Goudsmit et al.~revised MELD-Na using retrospective
data, leading to the \emph{ReMELD-Na} score.

In May 2023, the ELIAC recommended basing Eurotransplant liver allocation
on ReMELD-Na instead of UNOS-MELD. A matter of concern for ELIAC was
that ReMELD-Na scores range from 1 to 36, whereas UNOS-MELD scores range
from 6 to 40. Basing laboratory MELD scores on ReMELD-Na without
updating the MELD-curve could therefore mean that non-exception
candidates -- who depend on lab-MELD for access to transplantation -- could
only receive match-MELD scores up to 36, whereas candidates with (N)SE
or PED-MELD scores could still receive match-MELD scores up to 40. A
switch to ReMELD-Na could thereby inadvertently give extra priority to
candidates with exception points, who appear to be overprioritized already \citep{umgelterDisparitiesEurotransplantLiver2017a}.

To support this discussion, we simulated two scenarios for liver allocation using
ReMELD-Na. For both scenarios, lab-MELD scores were based on ReMELD-Na, calculated
using:

\[
\begin{aligned}
\text{ReMELD-Na} &= 7.85 + 9.03\ln(\text{crea}) + 2.97\ln(\text{bili}) + 9.52\ln(\text{INR}) + \\
& + 0.392(138.6 - \text{Na}) - 0.351(138.6 - \text{Na})\ln(\text{crea})
\end{aligned}
\]

In the first scenario, the exception score system remained unchanged,
which means that PED-MELD and (N)SE-MELD scores are calculated based on
the UNOS-MELD survival curve. This curve is given by the following
formula:

\[
S_{90}^{\texttt{UNOS-MELD}} = 0.98037^{0.17557(\texttt{UNOS-MELD} - 10)}.
\]

In the second scenario, PED-MELD and (N)SE-MELD scores are instead
calculated based on a survival curve developed specifically for
ReMELD-Na. This curve was obtained by estimating a Cox proportional hazards
model on Eurotransplant registry data with adjustment for ReMELD-Na.
The obtained ReMELD-Na curve is given by the following formula:

\[
S_{90}^{\text{ReMELD-Na}} = 0.9745^{0.2216(\text{ReMELD-Na} - 10)}
\]

Table \ref{tab:tab6} shows simulated waiting list outcomes under
these ReMELD-Na policies, summarized over 50 simulations. The
third column shows that introducing ReMELD-Na without updating the
S-curve results in approximately 70 extra waiting list deaths in total
(p\textless0.001), which corresponds to 15--20 extra waiting list deaths per
year. This finding is likely explained by ELIAC's concern that a switch
to ReMELD-Na inadvertently deprioritizes candidates who depend on
lab-MELD scores for access to transplantation. The fourth column shows
that switching to ReMELD-Na with the updated S-curve could have averted
25 waiting list deaths in total (p\textless0.001).

\begingroup
\sisetup{range-phrase = {\,\textemdash\,}}

\begin{table}[h]
\caption{Simulated number of waiting list exits for candidates between January 1, 2016, and January 1, 2020, under UNOS-MELD and ReMELD-Na, with and without updating the S-curve. The numbers in brackets are 95\% interquantile ranges. Paired t-tests were used to test whether either scenario significantly affected waiting list outcomes.}
\label{tab:tab6}
\centering
\resizebox{1\ifdim\width>\linewidth\linewidth\else\width\fi}{!}{%

\begin{tabular}{lccc}
    \toprule
    exit status & \makecell{UNOS-MELD\\(current)} & \makecell{ReMELD-Na\\(old S-curve)} & \makecell{ReMELD-Na\\(updated S-curve)}\\
    \midrule
    waiting list deaths & 1633.4 [1581.2-1685.6] & 1703.3*** [1643.2-1763.3] & 1608.3*** [1546.9-1669.7]\\
    transplantations & 6414.5 [6400.1-6428.9] & 6417.7* [6401.6-6433.7] & 6410.3*** [6396.7-6423.8]\\
    removed & 859.9 [834.8-885] & 837.9*** [812.9-862.8] & 870.4*** [842.7-898.1]\\
    \bottomrule
\end{tabular}
}
\parbox{\textwidth}{\footnotesize \smallskip $^{*}$p < 0.05; $^{**}$p < 0.01; $^{***}$p < 0.001}
\end{table}

\endgroup
\FloatBarrier

An important limitation of this analysis is that serum sodium
measurements are not available for most candidates in Eurotransplant
registry data. When serum sodium was missing in simulations, we
calculated ReMELD-Na scores with a serum sodium level of 138.6 mmol/l
(which results in 0 points being awarded for hyponatremia). Because this
means that hyponatremia could not be prioritized in all
candidates, the reduction of 25 waiting list deaths is a
conservative projection.

Whether serum sodium is reported to Eurotransplant depends foremostly on
the candidate's center of listing because centers either never or
almost always report serum sodium to Eurotransplant. This has motivated us to also assess
the number of waiting list deaths, stratified by the center of
listing's track record of reporting serum sodium. For this, we
categorize centers into centers with high, medium and low serum sodium
completeness. Table
\ref{tab:tab6} shows the total number of
waiting list deaths per type of center. It can be seen that the
reduction in waiting list deaths is indeed concentrated in centers with
high serum sodium completeness, with 28 fewer waiting list deaths observed
in such centers. Under the (optimistic) assumption that this reduction in
waiting list deaths is representative for all centers in Eurotransplant,
using ReMELD-Na as the basis for liver allocation could have prevented
up to 80 waiting list deaths over the simulation period.

\begin{table}[h]
\caption{Simulated number of waiting list deaths for candidates between January 1, 2016, and January 1, 2020, under UNOS-MELD (current) and ReMELD-Na with an updated S-curve.}
\label{tab:tab7}
\centering
\resizebox{1\ifdim\width>\linewidth\linewidth\else\width\fi}{!}{%

\begin{tabular}{C{.20\linewidth}C{.10\linewidth}C{.10\linewidth}C{.32\linewidth}C{.32\linewidth}}
\toprule
\makecell{serum sodium\\completeness} & \makecell{center\\count} & \makecell{patient\\count} & \makecell{waiting list deaths \\under UNOS-MELD} & \makecell{waiting list deaths \\under ReMELDna}\\
\midrule
high & 14 & 3995 & 589.3 [554.8-623.9] & 561*** [518.6-603.4]\\
medium & 13 & 3098 & 446.9 [404-489.7] & 446.9 [402.1-491.7]\\
low & 10 & 4775 & 590.3 [551.3-629.3] & 593.8 [555.3-632.4]\\
\bottomrule
\end{tabular}
}
\parbox{\textwidth}{\footnotesize \smallskip high: serum sodium known for >75\% of candidates; medium: serum sodium known for >50\% of candidates; low: otherwise.\\ $^{*}$ p < 0.05, $^{**}$ p < 0.01, $^{***}$ p < 0.001}
\end{table}

\FloatBarrier

\section{Conclusions and discussion}\label{sec:elasconclusion}

The Eurotransplant Liver Allocation System (ELAS) has changed little
since it was introduced in December 2006, despite the fact that several
areas of improvement have been identified by ELIAC. This lack of
development in liver allocation policies stands in contrast to other
organ allocation regions. For instance, in the United Kingdom the NHS
has replaced MELD-based liver allocation in 2018 by a benefits-based
allocation scheme \citep{Allen2024}, and liver allocation in the United
States was based on MELD-Na in 2016 and switched to MELD 3.0 in 2023
\citep{kimMELD3point0}. One major explanation for the lack of policy
development within Eurotransplant is that new allocation policies have
to be mutually agreed upon by all member countries. Aligning the
interests and perspectives of these member countries is challenging,
because the widely varying organ donation rates among ET member
countries mean that the countries face different challenges in liver
allocation.

To help align interests and perspectives of the Eurotransplant member
countries, Eurotransplant needs a tool which can simulate the impact of
proposed policy changes, as well as their unintended consequences. While
such tools are already routinely used for organ allocation policy
development in other geographic regions, the ELAS simulator is the first
simulator tailored to the unique challenges of Eurotransplant as a
multi-national organ exchange organization.

A key matter in the realization of the potential of the ELAS simulator
is its reputation, and the trust required by various stakeholders. To
build such trust in the ELAS simulator as a policy evaluation tool, we
have validated the ELAS simulator. We have
demonstrated that the ELAS simulator is able to closely replicate most
patterns in Eurotransplant liver allocation. In our own view and the
view of ELIAC, the remaining differences are sufficiently small for the
ELAS simulator to contribute to liver allocation policy evaluation
within Eurotransplant.

We have illustrated this with two case studies. The case study on the Belgian exception point system was requested by BeLIAC, and guided the decision to recommend a cap of 28 on (N)SE scores in Belgium.
The case study on ReMELD-Na was used to support a recommendation to switch allocation based on ReMELD-Na, which has happened on March 25, 2025.

We have illustrated this with two case studies. In the first case study,
we collaborated with the Belgian Liver and Intestine Advisory Committee
(BeLIAC) to assess how the Belgian exception system can be revised to
avoid that candidates who are ineligible for exception points are
crowded out of transplantation. A follow-up question from BeLIAC was
whether it would be possible to only curb MELD-SE scores for candidates
with hepatocellular carcinoma (HCC), who represent 70\% of the candidates
receiving (N)SEs. This topic is pending discussion in BeLIAC. For the
second case study, we have evaluated the impact of basing Eurotransplant
liver allocation on ReMELD-Na scores instead of UNOS-MELD scores.
Simulations for this case study suggested that a switch to ReMELD-Na can
indeed reduce waiting list mortality, but only if the exception scores
are appropriately re-scaled. Based on these results, the ELIAC has
recommended the Eurotransplant board to implement ReMELD-Na with
re-scaled exception points, and this proposal is scheduled for
implementation in March 2025.

This shows that the ELAS simulator has become useful for policy evaluation.
However, the software also has limitations. A first limitation is that our simulations were driven by
Eurotransplant registry data, which does not capture all information relevant
for liver allocation. For example, in the second case study we have seen that
serum sodium is not available in Eurotransplant registry data for many candidates, which
complicates simulating the impact of allocation principles that are
based on serum sodium. This limitation could be addressed by prospective
data collection of factors relevant to allocation. Secondly, statistical models
for graft offer acceptance behavior and post-transplant survival were calibrated
to historical data. New developments in liver transplantation threaten the external validity of
these models. An example of such a development is machine perfusion,
which is enhancing outcomes after liver transplantation
\citep{terraultLiverTransplantation20232023} and is altering graft offer
acceptance behavior of transplantation centers. This limitation can
potentially be addressed by recalibrating statistical models with
contemporary data.

From a policy point of view, there remain several areas in which
Eurotransplant's liver allocation system may be improved. These areas
include taking post-transplant outcomes into account when allocating
livers \citep{Allen2024}, broader geographic sharing of livers for candidates
with very high MELD scores
\citep{massieEarlyChangesLiver2015, Ravaioli2022}, and
rectifying sex disparity in liver waiting list outcomes
\citep{deFerranteSexDisparityLiver2024}. The ELAS simulator will continue to
play a role in informing the ELIAC of the impact of policy changes
related to these topics.

\section*{Acknowledgments}\label{acknowledgments}
\addcontentsline{toc}{section}{Acknowledgments}

We are grateful to prof. Xavier Verhelst, prof. Géraldine Dahlqvist, and
prof. Frederik Nevens for their advice for the BeLIAC case study. We
thank members of the ELIAC for the idea to conduct the ReMELD-Na case
study. We thank dr. Marko Boon for providing feedback on this
manuscript.

\begingroup
\hspace{\parindent}
\setlength{\parindent}{-0.25in}
\setlength{\leftskip}{0.25in}
\setlength{\parskip}{0pt}
\FloatBarrier
\newpage
\bibliography{thesis}

\begin{thebibliography}{40}
\providecommand{\natexlab}[1]{#1}
\providecommand{\url}[1]{#1}
\csname url@samestyle\endcsname
\providecommand{\newblock}{\relax}
\providecommand{\bibinfo}[2]{#2}
\providecommand{\BIBentrySTDinterwordspacing}{\spaceskip=0pt\relax}
\providecommand{\BIBentryALTinterwordstretchfactor}{4}
\providecommand{\BIBentryALTinterwordspacing}{\spaceskip=\fontdimen2\font plus
\BIBentryALTinterwordstretchfactor\fontdimen3\font minus
  \fontdimen4\font\relax}
\providecommand{\BIBforeignlanguage}[2]{{%
\expandafter\ifx\csname l@#1\endcsname\relax
\typeout{** WARNING: IEEEtranN.bst: No hyphenation pattern has been}%
\typeout{** loaded for the language `#1'. Using the pattern for}%
\typeout{** the default language instead.}%
\else
\language=\csname l@#1\endcsname
\fi
#2}}
\providecommand{\BIBdecl}{\relax}
\BIBdecl

\bibitem[Umgelter et~al.(2017)Umgelter, Hapfelmeier, Kopp, {van Rosmalen},
  Rogiers, Guba, et~al.]{umgelterDisparitiesEurotransplantLiver2017a}
A.~Umgelter, A.~Hapfelmeier, W.~Kopp, M.~{van Rosmalen}, X.~Rogiers, M.~Guba
  \emph{et~al.}, ``\BIBforeignlanguage{en}{Disparities in {{Eurotransplant}}
  liver transplantation wait-list outcome between patients with and without
  {{Model for End-Stage Liver Disease}} exceptions},''
  \emph{\BIBforeignlanguage{en}{Liver Transplantation}}, vol.~23, no.~10, pp.
  1256--1265, 2017.

\bibitem[Kamath et~al.(2001)Kamath, Wiesner, Malinchoc, Kremers, Therneau,
  Kosberg, et~al.]{kamathModelPredictSurvival2001}
P.~S. Kamath, R.~H. Wiesner, M.~Malinchoc, W.~Kremers, T.~M. Therneau, C.~L.
  Kosberg \emph{et~al.}, ``A model to predict survival in patients with
  end-stage liver disease,'' \emph{Hepatology}, vol.~33, no.~2, pp. 464--470,
  2001.

\bibitem[Moylan et~al.(2008)Moylan, Brady, Johnson, Smith, Tuttle-Newhall, and
  Muir]{moylanDisparitiesLiverTransplantation2008}
C.~A. Moylan, C.~W. Brady, J.~L. Johnson, A.~D. Smith, J.~E. Tuttle-Newhall,
  and A.~J. Muir, ``\BIBforeignlanguage{eng}{Disparities in liver
  transplantation before and after introduction of the {{MELD}} score},''
  \emph{\BIBforeignlanguage{eng}{JAMA}}, vol. 300, no.~20, pp. 2371--2378,
  2008.

\bibitem[{de Ferrante} et~al.(2025){de Ferrante}, de~Rosner-van Rosmalen,
  Smeulders, Vogelaar, and Spieksma]{deFerranteSexDisparityLiver2024}
H.~C. {de Ferrante}, M.~de~Rosner-van Rosmalen, B.~M.~L. Smeulders,
  S.~Vogelaar, and F.~C.~R. Spieksma, ``\BIBforeignlanguage{English}{Sex
  disparity in liver allocation within {{Eurotransplant}}},''
  \emph{\BIBforeignlanguage{English}{American Journal of Transplantation}},
  vol.~25, no.~1, pp. 139--149, 2025.

\bibitem[Pritsker et~al.(1995)Pritsker, Kuhl, Roberts, Allen, Burdick, Martin,
  et~al.]{Pritsker1995}
A.~A.~B. Pritsker, M.~E. Kuhl, J.~P. Roberts, M.~D. Allen, J.~F. Burdick, D.~L.
  Martin \emph{et~al.}, ``Organ transplantation policy evaluation,'' in
  \emph{Proceedings of the 27th Conference on Winter simulation - WSC '95},
  ser. WSC ’95, C.~Alexopoulos, K.~Kang, W.~R. Lilegdon, and D.~Goldsman,
  Eds.\hskip 1em plus 0.5em minus 0.4em\relax INFORMS, 1995, pp. 1314--1323.

\bibitem[Wujciak and Opelz(1993)]{wujciakProposalImprovedCadaver1993a}
T.~Wujciak and G.~Opelz, ``\BIBforeignlanguage{en}{A proposal for improved
  cadaver kidney allocation},''
  \emph{\BIBforeignlanguage{en}{Transplantation}}, vol.~56, no.~6, pp.
  1513--1517, 1993.

\bibitem[{de Meester} et~al.(1998){de Meester}, Persijn, Wujciak, Opelz, and
  Vanrenterghem]{demeesterNewEurotransplantKidney1998}
J.~{de Meester}, G.~G. Persijn, T.~Wujciak, G.~Opelz, and Y.~Vanrenterghem,
  ``\BIBforeignlanguage{eng}{The new {{Eurotransplant}} kidney allocation
  system: report one year after implementation.}''
  \emph{\BIBforeignlanguage{eng}{Transplantation}}, vol.~66, no.~9, pp.
  1154--1159, 1998.

\bibitem[Thompson et~al.(2004)Thompson, Waisanen, Wolfe, Merion, McCullough,
  and Rodgers]{ThompsonXSAM2004}
D.~Thompson, L.~Waisanen, R.~Wolfe, R.~M. Merion, K.~McCullough, and
  A.~Rodgers, ``\BIBforeignlanguage{en}{Simulating the allocation of organs for
  transplantation},'' \emph{\BIBforeignlanguage{en}{Health Care Management
  Science}}, vol.~7, no.~4, pp. 331--338, 2004.

\bibitem[Kim et~al.(2008)Kim, Biggins, Kremers, Wiesner, Kamath, Benson,
  et~al.]{kimHyponatremiaMortalityPatients2008a}
W.~R. Kim, S.~W. Biggins, W.~K. Kremers, R.~H. Wiesner, P.~S. Kamath, J.~T.
  Benson \emph{et~al.}, ``\BIBforeignlanguage{eng}{Hyponatremia and mortality
  among patients on the liver-transplant waiting list},''
  \emph{\BIBforeignlanguage{eng}{The New England Journal of Medicine}}, vol.
  359, no.~10, pp. 1018--1026, 2008.

\bibitem[Kim et~al.(2021)Kim, Mannalithara, Heimbach, Kamath, Asrani, Biggins,
  et~al.]{kimMELD3point0}
W.~R. Kim, A.~Mannalithara, J.~K. Heimbach, P.~S. Kamath, S.~K. Asrani, S.~W.
  Biggins \emph{et~al.}, ``{{MELD}} 3.0: the {{Model for End-Stage Liver
  Disease}} updated for the modern era,'' \emph{Gastroenterology}, vol. 161,
  no.~6, pp. 1887--1895.e4, 2021.

\bibitem[Freeman et~al.(2004)Freeman, Wiesner, Roberts, McDiarmid, Dykstra, and
  Merion]{freemanImprovingLiverAllocation2004}
R.~B. Freeman, R.~H. Wiesner, J.~P. Roberts, S.~McDiarmid, D.~M. Dykstra, and
  R.~M. Merion, ``Improving liver allocation: {{MELD}} and {{PELD}},''
  \emph{American Journal of Transplantation}, vol.~4, pp. 114--131, 2004.

\bibitem[Axelrod et~al.(2011)Axelrod, Gheorghian, Schnitzler, Dzebisashvili,
  Salvalaggio, Tuttle-Newhall, et~al.]{axelrodEconomicImplicationsBroader2011}
D.~A. Axelrod, A.~Gheorghian, M.~A. Schnitzler, N.~Dzebisashvili, P.~R.
  Salvalaggio, J.~Tuttle-Newhall \emph{et~al.}, ``The economic implications of
  broader sharing of liver allografts,'' \emph{American Journal of
  Transplantation}, vol.~11, no.~4, pp. 798--807, 2011.

\bibitem[Gentry et~al.(2013)Gentry, Massie, Cheek, Lentine, Chow, Wickliffe,
  et~al.]{gentryAddressingGeographicDisparities2013}
S.~E. Gentry, A.~B. Massie, S.~W. Cheek, K.~L. Lentine, E.~H. Chow, C.~E.
  Wickliffe \emph{et~al.}, ``Addressing geographic disparities in liver
  transplantation through redistricting,'' \emph{American Journal of
  Transplantation}, vol.~13, no.~8, pp. 2052--2058, 2013.

\bibitem[Goel et~al.(2018)Goel, Kim, Pyke, Schladt, Kasiske, Snyder,
  et~al.]{goelLiverSimulatedAllocation2018}
A.~Goel, W.~R. Kim, J.~Pyke, D.~P. Schladt, B.~L. Kasiske, J.~J. Snyder
  \emph{et~al.}, ``\BIBforeignlanguage{en}{{Liver Simulated Allocation
  Modeling}: were the predictions accurate for {{Share 35}}?}''
  \emph{\BIBforeignlanguage{en}{Transplantation}}, vol. 102, no.~5, pp.
  769--774, 2018.

\bibitem[Akshat et~al.(2024)Akshat, Gentry, and
  Raghavan]{akshatHeterogeneousDonorCircles2024}
S.~Akshat, S.~E. Gentry, and S.~Raghavan,
  ``\BIBforeignlanguage{en}{Heterogeneous donor circles for fair liver
  transplant allocation},'' \emph{\BIBforeignlanguage{en}{Health Care
  Management Science}}, vol.~27, no.~1, pp. 20--45, 2024.

\bibitem[Perito et~al.(2019)Perito, Mogul, VanDerwerken, Mazariegos, Bucuvalas,
  Book, et~al.]{peritoImpactIncreasedAllocation2019}
E.~R. Perito, D.~B. Mogul, D.~VanDerwerken, G.~Mazariegos, J.~Bucuvalas,
  L.~Book \emph{et~al.}, ``The impact of increased allocation priority for
  children awaiting liver transplant: a liver simulated allocation model
  {{(LSAM)}} analysis,'' \emph{Journal of Pediatric Gastroenterology and
  Nutrition}, vol.~68, no.~4, pp. 472--479, 2019.

\bibitem[Heimbach et~al.(2015)Heimbach, Hirose, Stock, Schladt, Xiong, Liu,
  et~al.]{heimbachDelayedHepatocellularCarcinoma2015}
J.~K. Heimbach, R.~Hirose, P.~G. Stock, D.~P. Schladt, H.~Xiong, J.~Liu
  \emph{et~al.}, ``\BIBforeignlanguage{eng}{Delayed hepatocellular carcinoma
  {{Model for End-Stage Liver Disease}} exception score improves disparity in
  access to liver transplant in the {{United States}}},''
  \emph{\BIBforeignlanguage{eng}{Hepatology}}, vol.~61, no.~5, pp. 1643--1650,
  2015.

\bibitem[Bernards et~al.(2022)Bernards, Lee, Leung, Akan, Gan, Zhao,
  et~al.]{bernardsAwardingAdditionalMELD2022}
S.~Bernards, E.~Lee, N.~Leung, M.~Akan, K.~Gan, H.~Zhao \emph{et~al.},
  ``\BIBforeignlanguage{eng}{Awarding additional {{MELD}} points to the
  shortest waitlist candidates improves sex disparity in access to liver
  transplant in the {{United States}}},''
  \emph{\BIBforeignlanguage{eng}{American Journal of Transplantation}},
  vol.~22, no.~12, pp. 2912--2920, 2022.

\bibitem[Jacquelinet et~al.(2006)Jacquelinet, Audry, Golbreich, Antoine,
  Rebibou, Claquin3, et~al.]{jacquelinet2006changing}
C.~Jacquelinet, B.~Audry, C.~Golbreich, C.~Antoine, J.-M. Rebibou, J.~Claquin3
  \emph{et~al.}, ``Changing kidney allocation policy in {{France:}} the value
  of simulation,'' \emph{AMIA F Symposium Proceedings}, vol. 2006, pp.
  374--378, 2006.

\bibitem[Bayer et~al.(2021)Bayer, Audry, Antoine, Jasseron, Legeai, Bastien,
  et~al.]{bayer2021removing}
F.~Bayer, B.~Audry, C.~Antoine, C.~Jasseron, C.~Legeai, O.~Bastien
  \emph{et~al.}, ``Removing administrative boundaries using a gravity model for
  a national liver allocation system,'' \emph{American Journal of
  Transplantation}, vol.~21, no.~3, pp. 1080--1091, 2021.

\bibitem[Watson et~al.(2020)Watson, Johnson, and Mumford]{watson2020overview}
C.~J.~E. Watson, R.~J. Johnson, and L.~Mumford,
  ``\BIBforeignlanguage{en}{Overview of the evolution of the {{UK}} kidney
  allocation schemes},'' \emph{\BIBforeignlanguage{en}{Current Transplantation
  Reports}}, vol.~7, no.~2, pp. 140--144, 2020.

\bibitem[{de Klerk} et~al.(2021){de Klerk}, van Gestel, {van de Wetering}, Kho,
  de~Sterke, Betjes, et~al.]{deKlerk2021creating}
M.~{de Klerk}, J.~A.~K. van Gestel, J.~{van de Wetering}, M.~L. Kho, S.~M.
  de~Sterke, M.~G.~H. Betjes \emph{et~al.}, ``\BIBforeignlanguage{eng}{Creating
  options for difficult-to-match kidney transplant candidates},''
  \emph{\BIBforeignlanguage{eng}{Transplantation}}, vol. 105, no.~1, pp.
  240--248, 2021.

\bibitem[Shoaib et~al.(2022)Shoaib, Prabhakar, Mahlawat, and
  Ramamohan]{shoaibDiscreteeventSimulationModel2022}
M.~Shoaib, U.~Prabhakar, S.~Mahlawat, and V.~Ramamohan,
  ``\BIBforeignlanguage{en}{A discrete-event simulation model of the kidney
  transplantation system in rajasthan, india},''
  \emph{\BIBforeignlanguage{en}{Health Systems}}, vol.~11, no.~1, pp. 30--47,
  2022.

\bibitem[Bertsimas et~al.(2020)Bertsimas, Papalexopoulos, Trichakis, Wang,
  Hirose, and Vagefi]{bertsimasBalancingEfficiencyFairness2020b}
D.~Bertsimas, T.~Papalexopoulos, N.~Trichakis, Y.~Wang, R.~Hirose, and P.~A.
  Vagefi, ``\BIBforeignlanguage{en}{Balancing efficiency and fairness in liver
  transplant access: tradeoff curves for the assessment of organ distribution
  policies},'' \emph{\BIBforeignlanguage{en}{Transplantation}}, vol. 104,
  no.~5, pp. 981--987, 2020.

\bibitem[Papalexopoulos et~al.(2024)Papalexopoulos, Alcorn, Bertsimas, Goff,
  Stewart, and Trichakis]{papalexopoulosReshapingNationalOrgan2024}
T.~Papalexopoulos, J.~Alcorn, D.~Bertsimas, R.~Goff, D.~Stewart, and
  N.~Trichakis, ``Reshaping national organ allocation policy,''
  \emph{Operations Research}, vol.~72, no.~4, pp. 1475--1486, 2024.

\bibitem[Mankowski et~al.(2023)Mankowski, Wood, Segev, and
  Gentry]{mankowskiRemovingGeographicBoundaries2023}
M.~A. Mankowski, N.~L. Wood, D.~L. Segev, and S.~E. Gentry, ``Removing
  geographic boundaries from liver allocation: a method for designing
  continuous distribution scores,'' \emph{Clinical Transplantation}, vol.~37,
  no.~9, p. e15017, 2023.

\bibitem[{Eurotransplant}(2024)]{et_donor_pmp}
\BIBentryALTinterwordspacing
{Eurotransplant}, ``{{Eurotransplant}} {{Statistics Library}}. {{Report}}
  1031p: deceased donors used, per million population, by year, by donor
  country,'' Online, 2024. [Online]. Available:
  \url{https://statistics.eurotransplant.org/reportloader.php?report=10867-33157&format=html&download=0}
\BIBentrySTDinterwordspacing

\bibitem[Jochmans et~al.(2017)Jochmans, {van Rosmalen}, Pirenne, and
  Samuel]{jochmansAdultLiverAllocation2017}
I.~Jochmans, M.~{van Rosmalen}, J.~Pirenne, and U.~Samuel,
  ``\BIBforeignlanguage{eng}{Adult liver allocation in {{Eurotransplant}}},''
  \emph{\BIBforeignlanguage{eng}{Transplantation}}, vol. 101, no.~7, pp.
  1542--1550, 2017.

\bibitem[{Eurotransplant}()]{ETLiverMan2025}
\BIBentryALTinterwordspacing
{Eurotransplant}, ``{Eurotransplant Manual} manual. chapter 4: {ET Liver
  Allocation System (ELAS)}.'' [Online]. Available:
  \url{https://www.eurotransplant.org/allocation/eurotransplant-manual/}
\BIBentrySTDinterwordspacing

\bibitem[Shechter et~al.(2005)Shechter, Bryce, Alagoz, Kreke, Stahl, Schaefer,
  et~al.]{shechterClinicallyBasedDiscreteEvent2005}
S.~M. Shechter, C.~L. Bryce, O.~Alagoz, J.~E. Kreke, J.~E. Stahl, A.~J.
  Schaefer \emph{et~al.}, ``\BIBforeignlanguage{en}{A clinically based
  discrete-event simulation of end-stage liver disease and the organ allocation
  process},'' \emph{\BIBforeignlanguage{en}{Medical Decision Making}}, vol.~25,
  no.~2, pp. 199--209, 2005.

\bibitem[Ratcliffe et~al.(2001)Ratcliffe, Young, Buxton, Eldabi, Paul,
  Burroughs, et~al.]{ratcliffeSimulationModellingApproach2001}
J.~Ratcliffe, T.~Young, M.~Buxton, T.~Eldabi, R.~Paul, A.~Burroughs
  \emph{et~al.}, ``\BIBforeignlanguage{eng}{A simulation modelling approach to
  evaluating alternative policies for the management of the waiting list for
  liver transplantation},'' \emph{\BIBforeignlanguage{eng}{Health Care
  Management Science}}, vol.~4, no.~2, pp. 117--124, 2001.

\bibitem[{Scientific Registry of Transplant Recipients}(2019)]{SRTR2019}
\BIBentryALTinterwordspacing
{Scientific Registry of Transplant Recipients},
  ``\BIBforeignlanguage{en}{{Liver Simulation Allocation Model} - user's
  guide},'' 2019. [Online]. Available:
  \url{https://www.srtr.org/media/1361/lsam-2019-User-Guide.pdf}
\BIBentrySTDinterwordspacing

\bibitem[Wood et~al.(2021)Wood, Mogul, Perito, VanDerwerken, Mazariegos, Hsu,
  et~al.]{WoodPEDLSAM2021}
N.~L. Wood, D.~B. Mogul, E.~R. Perito, D.~VanDerwerken, G.~V. Mazariegos, E.~K.
  Hsu \emph{et~al.}, ``{Liver Simulated Allocation Model} does not effectively
  predict organ offer decisions for pediatric liver transplant candidates,''
  \emph{American Journal of Transplantation}, vol.~21, no.~9, pp. 3157--3162,
  2021.

\bibitem[Agarwal et~al.(2021)Agarwal, Ashlagi, Rees, Somaini, and
  Waldinger]{agarwalEquilibriumAllocationsAlternative2021}
N.~Agarwal, I.~Ashlagi, M.~A. Rees, P.~Somaini, and D.~Waldinger,
  ``\BIBforeignlanguage{en}{Equilibrium allocations under alternative waitlist
  designs: evidence from deceased donor kidneys},''
  \emph{\BIBforeignlanguage{en}{Econometrica}}, vol.~89, no.~1, pp. 37--76,
  2021.

\bibitem[Carson(1989)]{carsonVerificationValidationConsultants1989}
J.~S. Carson, ``Verification and validation: a consultant's perspective,'' in
  \emph{Proceedings of the 21st Conference on {Winter} simulation}, ser. {WSC}
  '89, New York, NY, USA, 1989, pp. 552--558.

\bibitem[Law(2015)]{lawSimulationModelingAnalysis2015}
A.~M. Law, ``\BIBforeignlanguage{en}{Variance-reduction techniques},'' in
  \emph{\BIBforeignlanguage{en}{Simulation {Modeling} and {Analysis}}}, fifth
  edition~ed., ser. {McGraw}-{Hill} series in industrial engineering and
  management science.\hskip 1em plus 0.5em minus 0.4em\relax Dubuque:
  McGraw-Hill Education, 2015, ch.~11, pp. 588--596.

\bibitem[Allen et~al.(2024)Allen, Taylor, Gimson, and Thorburn]{Allen2024}
E.~Allen, R.~Taylor, A.~Gimson, and D.~Thorburn,
  ``\BIBforeignlanguage{English}{Transplant benefit-based offering of deceased
  donor livers in the {{United Kingdom}}},''
  \emph{\BIBforeignlanguage{English}{Journal of Hepatology}}, vol.~81, no.~3,
  pp. 471--478, 2024.

\bibitem[Terrault et~al.(2023)Terrault, Francoz, Berenguer, Charlton, and
  Heimbach]{terraultLiverTransplantation20232023}
N.~A. Terrault, C.~Francoz, M.~Berenguer, M.~Charlton, and J.~Heimbach,
  ``\BIBforeignlanguage{English}{Liver transplantation 2023: status report,
  current and future challenges},'' \emph{\BIBforeignlanguage{English}{Clinical
  Gastroenterology and Hepatology}}, vol.~21, no.~8, pp. 2150--2166, 2023.

\bibitem[Massie et~al.(2015)Massie, Chow, Wickliffe, Luo, Gentry, Mulligan,
  et~al.]{massieEarlyChangesLiver2015}
A.~B. Massie, E.~K.~H. Chow, C.~E. Wickliffe, X.~Luo, S.~E. Gentry, D.~C.
  Mulligan \emph{et~al.}, ``\BIBforeignlanguage{en}{Early changes in liver
  distribution following implementation of {{Share 35}}},''
  \emph{\BIBforeignlanguage{en}{American Journal of Transplantation}}, vol.~15,
  no.~3, pp. 659--667, 2015.

\bibitem[Ravaioli et~al.(2022)Ravaioli, Lai, Sessa, Ghinolfi, Fallani, Patrono,
  et~al.]{Ravaioli2022}
M.~Ravaioli, Q.~Lai, M.~Sessa, D.~Ghinolfi, G.~Fallani, D.~Patrono
  \emph{et~al.}, ``Impact of {{MELD}} 30-allocation policy on liver transplant
  outcomes in {{Italy}},'' \emph{Journal of Hepatology}, vol.~76, no.~3, pp.
  619--627, 2022.

\end{thebibliography}
\newpage
\endgroup

\appendix

\section{Supplementary materials}\label{chap:supp}

\begin{algorithm}[ht!]
\caption{Pseudocode for the liver graft offering module}
\label{alg:graft-offering}
\begin{algorithmic}[1]
\Require 
donor \texttt{d}; ordered match list of candidates \texttt{M} = $(c_1, c_2, ..., c_n)$; threshold for non-standard allocation \texttt{k}; boolean flag \texttt{forcePlacement}.
\Ensure 
Returns either a candidate who accepts the graft or signals a discard.

\vspace{6pt}
\Function{GraftOffering}{\texttt{d, M, k, forcePlacement}}
  \State \texttt{r} $\gets$ 0 \Comment rejection counter
  \State \texttt{centerDeclined} $\gets \emptyset$
  \State \texttt{rescueMode} $\gets$ \texttt{False}

  \ForAll{\texttt{c} \textbf{in} \texttt{M}}
    \If{\texttt{r} $\ge$ \texttt{k} \textbf{and not} \texttt{rescueMode}}
      \State \texttt{rescueMode} $\gets$ \texttt{True}
      \State \Call{ReorderToNonStandard}{\texttt{M}}
    \EndIf

    \If{\texttt{c.center} $\in$ \texttt{centerDeclined}}
      \State \textbf{continue}
    \EndIf

    \State $\alpha_c \gets$ \Call{LogisticCenterAcceptance}{\texttt{d, c}}
    \If{$\alpha_c < \texttt{Uniform(0,1)}$}
      \State \texttt{r} $\gets$ \texttt{r + 1}
      \State \texttt{centerDeclined} $\gets$ \texttt{centerDeclined} $\cup$ \{\texttt{c.center}\}
    \EndIf

    \If{\Call{ProfileCompatible}{\texttt{d, c, \texttt{rescueMode}}}}
      \State $\alpha_p \gets$ \Call{LogisticPatientAcceptance}{\texttt{d, c}}
      \If{$\alpha_p \ge \texttt{Uniform(0,1)}$}
        \State \textbf{return} \texttt{c}
      \Else
        \State \texttt{r} $\gets$ \texttt{r + 1}
      \EndIf
    \EndIf
  \EndFor

  \If{\texttt{forcePlacement} \textbf{is True} \textbf{and} \texttt{M} $\neq \emptyset$}
    \State \texttt{c*} $\gets \arg\max_{\texttt{c} \in \texttt{M}}$ \Call{LogisticPatientAcceptance}{\texttt{d, c}}
    \State \textbf{return} \texttt{c*}
  \Else
    \State \textbf{return} \textsc{Discard}
  \EndIf
\EndFunction

\vspace{6pt}
\Function{ProfileCompatible}{\texttt{d, c, \texttt{rescueMode}}}
  \State \textbf{return} \texttt{True} if \texttt{c} is HU/ACO or profile deems \texttt{d} acceptable; \texttt{False} otherwise
\EndFunction

\vspace{6pt}
\Function{LogisticCenterAcceptance}{\texttt{d, c}}
  \State \textbf{return} predicted acceptance probability from center-level logistic model
\EndFunction

\vspace{6pt}
\Function{LogisticPatientAcceptance}{\texttt{d, c}}
  \State \textbf{return} predicted acceptance probability from patient-level logistic model
\EndFunction

\end{algorithmic}
\end{algorithm}

\renewcommand\thefigure{\thesection\arabic{figure}}
\renewcommand\thetable{\thesection\arabic{table}}
\setcounter{figure}{0}
\setcounter{table}{0}

\FloatBarrier
\vspace{.1cm}
\newpage

\hypersetup{
  colorlinks = true,      
  linkcolor = blue,       
  urlcolor  = blue,       
  citecolor = blue,       
  pdfborder = {0 0 1}
}
\begingroup\fontsize{8}{10}\selectfont
\rowcolors{2}{gray!10}{white}
\begin{longtable}[t]{>{\raggedright\arraybackslash}p{0.16\textwidth} >{\raggedright\arraybackslash}p{0.50\textwidth} >{\raggedright\arraybackslash}p{0.32\textwidth}}
\caption{\label{tab:stab2}Main input data and parameters for the ELAS simulator. Columns list each input or parameter set, describe its role in the simulator, and provide typical examples or references.}\\
\toprule
\textbf{Input or parameter} & \textbf{Usage in the simulator} & \textbf{Example file or source} \\
\midrule
\endfirsthead

\caption[]{Main input data and parameters for the ELAS simulator \textit{(continued)}}\\
\toprule
\textbf{Input or Parameter} & \textbf{Usage in the simulator} & \textbf{Example file or source} \\
\midrule
\endhead

\midrule
\multicolumn{3}{r@{}}{\textit{(Continued on next page...)}} \\
\endfoot

\bottomrule
\endlastfoot

Simulation settings & 
Global parameters: simulation start and end dates, MELD formula, S-curve for exceptions, random seeds, paths to input streams. & 
YAML files — see \href{https://github.com/hansdeferrante/Eurotransplant_ELAS_simulator/tree/main/simulator/sim_yamls/templates}{Github example} \\

Obligation input stream & 
Specifies transplantations for which international sharing was mandatory (HU/ACO)
as well as when such obligations were fulfilled. This is an optional input stream
that can initialize the obligation system. & 
See \texttt{fake\_obligations.csv} as an example dataset on Github \\

Donor input stream & 
Defines each donated liver, including the donor's characteristics needed for allocation, prediction of rescue probabilities, center offer acceptance behavior, and post-transplant survival. & 
Eurotransplant registry or synthetic data — see \texttt{fake\_donors.csv} on \href{https://github.com/hansdeferrante/Eurotransplant_ELAS_simulator}{Github} \\

Candidate input stream & 
Lists each patient's static information (e.g., date of birth, blood group, listing center, time of listing). Also includes predictors for offer acceptance and post-transplant survival. & 
Eurotransplant registry or synthetic data — see \texttt{fake\_recipients.csv} on Github \\

Candidate status updates & 
Changes to MELD lab values, transitions to HU/ACO, SE/NSE/PED-MELD statuses, exit 
statuses. Includes a reporting time, which is measured relative to the corresponding
candidate's listing time. & 
Registry data with imputation or synthetic — see \texttt{fake\_patstat1.csv} on Github \\

Exception definitions & 
Contains SE, NSE, and PED rules per country, including their initial equivalent (e.g., 15\%),
increment frequency (e.g., every 90 days), and increment values (e.g., 10 percentage points). & 
These are described in the Eurotransplant liver manual \citep{ETLiverMan2025}.
Defaults are in \texttt{se\_rules.csv} on Github; 
example adaptations are also available such as \texttt{BELIAC\_se\_rules\_capped\_25.csv}\\

Non-standard allocation parameters & 
Cox model coefficients and baseline hazards used to predict the number of offers
made until rescue allocation is triggered. & 
See \texttt{coefs\_rescue\_triggered.csv} and \texttt{bh\_rescue\_triggered.csv} on \href{https://github.com/hansdeferrante/Eurotransplant_ELAS_simulator/tree/main/simulator/magic_values}{Github},
respectively.\\

Offer acceptance parameters & 
Logistic regression coefficients (odds ratios) for modeling likelihood of graft acceptance. & 
Separate for pediatric/adult and HU/ACO/elective — e.g., \texttt{coefs\_adult\_reg.csv} 
for adult, elective candidates on \href{https://github.com/hansdeferrante/Eurotransplant_ELAS_simulator/blob/main/simulator/magic_values/acceptance/coefs_adult_reg.csv}{Github}.\\

Post-transplant survival parameters & 
Weibull parameters for graft survival, and Kaplan-Meier estimates to simulate re-listing. & 
Estimated from 2012–2019 data — see \href{https://github.com/hansdeferrante/Eurotransplant_ELAS_simulator/tree/main/simulator/magic_values/post_txp}{Github folder} \\

\end{longtable}
\endgroup

\FloatBarrier
\vspace{.1cm}
\newpage

\begin{figure}[h]

{\centering \includegraphics[width=0.9\linewidth]{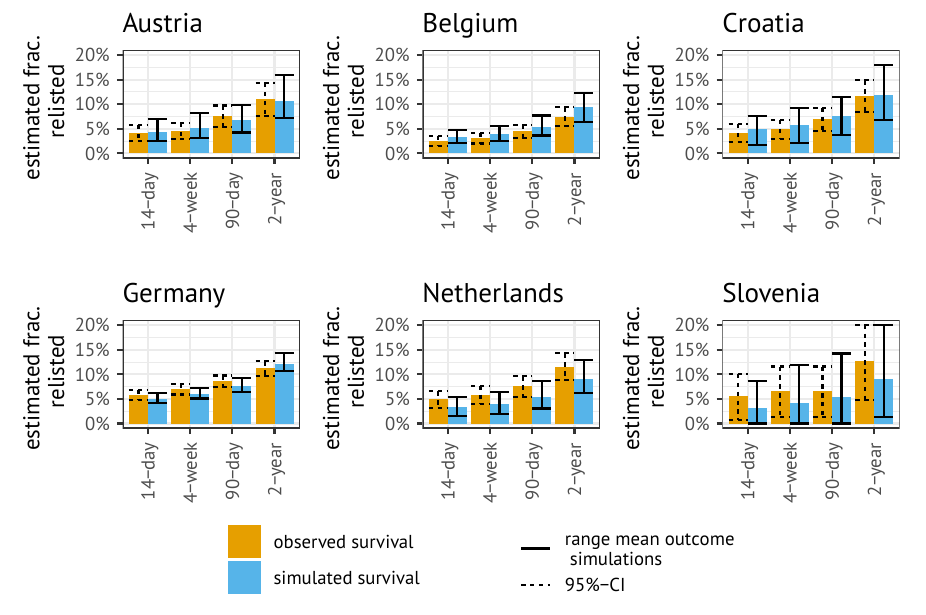} 

}

\caption{Estimated re-listing rates for transplant recipients, per country for different time horizons. Relisting rates were estimated with the Kaplan-Meier estimator, using time since transplantation as the timescale.}\label{fig:sfig1}
\end{figure}

\FloatBarrier
\vspace{.1cm}
\newpage

\section{Overview of 2016 (N)SE allocation rules in Eurotransplant}\label{app:nses}

\renewcommand{\thetable}{A\arabic{table}}
Table \ref{tab:nse_table} gives an overview of the (N)SEs and PED-MELD as used in ELAS. For a few SEs (e.g., primary hyperoxaluria), the awarded initial mortality equivalent depends on the SE criteria fulfilled by the candidate. For such SEs, the most commonly awarded initial mortality equivalents are shown.

\setlength\LTleft{-8mm}
\begingroup\fontsize{8}{10}\selectfont

\begin{longtable}[t]{>{\raggedright\arraybackslash}p{3cm}l>{\raggedright\arraybackslash}p{0.9cm}>{\raggedright\arraybackslash}p{0.9cm}>{\raggedright\arraybackslash}p{0.9cm}>{\raggedright\arraybackslash}p{0.9cm}>{\raggedright\arraybackslash}p{0.9cm}>{\raggedright\arraybackslash}p{0.9cm}>{\raggedright\arraybackslash}p{0.9cm}>{\raggedright\arraybackslash}p{0.9cm}}
\caption{\label{tab:nse_table}Overview of the (N)SE allocation rules used by Eurotransplant}\\
\toprule
\rowcolor{white}
Exception & Type & $($N$)$SE ID & Country & Initial equiv. (\%) & Upgrade incr. (\%) & Upgrade time (days) & Bonus amount (\%) & Max age & Next (N)SE\\
\rowcolor{gray!6}
\midrule
\endfirsthead
\caption[]{Overview of the (N)SE allocation rules used by Eurotransplant \textit{(continued)}}\\
\toprule
\rowcolor{white}
Exception & Type & $($N$)$SE ID & Country & Initial equiv. (\%) & Upgrade incr. (\%) & Upgrade time (days) & Bonus amount (\%) & Max age & Next (N)SE\\
\rowcolor{gray!6}
\midrule
\endhead
\midrule
\multicolumn{10}{r@{}}{\textit{(Continued on Next Page...)}}\
\endfoot
\bottomrule
\endlastfoot
\rowcolor{gray!6}
 &  & 1 & AU &  &  &  &  &  & \\
\rowcolor{gray!6}
\nopagebreak
 &  & 21 & SL &  &  &  &  &  & \\
\rowcolor{gray!6}
\nopagebreak
 &  & 81 & DE &  &  &  &  &  & \\
\rowcolor{gray!6}
\nopagebreak
\multirow{-4}{3cm}{\raggedright\arraybackslash Cholangiocarcinoma} & \multirow{-4}{*}{\raggedright\arraybackslash SE} & 101 & HR & \multirow{-4}{0.9cm}{\raggedright\arraybackslash 10} & \multirow{-4}{0.9cm}{\raggedright\arraybackslash 10} & \multirow{-4}{0.9cm}{\raggedright\arraybackslash 80} & \multirow{-4}{0.9cm}{\raggedright\arraybackslash } & \multirow{-4}{0.9cm}{\raggedright\arraybackslash } & \multirow{-4}{0.9cm}{\raggedright\arraybackslash }\\
\cmidrule{1-10}\pagebreak[0]
 &  & 2 & AU &  &  &  &  &  & \\
\nopagebreak
 &  & 22 & SL &  &  &  &  &  & \\
\nopagebreak
 &  & 51 & BE &  &  &  &  &  & \\
\nopagebreak
 &  & 82 & DE &  &  &  &  &  & \\
\nopagebreak
\multirow{-5}{3cm}{\raggedright\arraybackslash Cystic fibrosis} & \multirow{-5}{*}{\raggedright\arraybackslash SE} & 102 & HR & \multirow{-5}{0.9cm}{\raggedright\arraybackslash 10} & \multirow{-5}{0.9cm}{\raggedright\arraybackslash 10} & \multirow{-5}{0.9cm}{\raggedright\arraybackslash 80} & \multirow{-5}{0.9cm}{\raggedright\arraybackslash } & \multirow{-5}{0.9cm}{\raggedright\arraybackslash } & \multirow{-5}{0.9cm}{\raggedright\arraybackslash }\\
\cmidrule{1-10}\pagebreak[0]
\rowcolor{gray!6}
Familial Amyloidotic Polyneuropathy  & SE & 41 & NL & 10 & 10 & 80 &  &  & \\
\cmidrule{1-10}\pagebreak[0]
 &  & 3 & AU &  &  &  &  &  & \\
\nopagebreak
 &  & 23 & SL &  &  &  &  &  & \\
\nopagebreak
 &  & 52 & BE &  &  &  &  &  & \\
\nopagebreak
 &  & 83 & DE &  &  &  &  &  & \\
\nopagebreak
\multirow{-5}{3cm}{\raggedright\arraybackslash Familial Amyloidotic Polyneuropathy } & \multirow{-5}{*}{\raggedright\arraybackslash SE} & 103 & HR & \multirow{-5}{0.9cm}{\raggedright\arraybackslash 15} & \multirow{-5}{0.9cm}{\raggedright\arraybackslash 10} & \multirow{-5}{0.9cm}{\raggedright\arraybackslash 80} & \multirow{-5}{0.9cm}{\raggedright\arraybackslash } & \multirow{-5}{0.9cm}{\raggedright\arraybackslash } & \multirow{-5}{0.9cm}{\raggedright\arraybackslash }\\
\cmidrule{1-10}\pagebreak[0]
\rowcolor{gray!6}
Hepato-pulmonary syndrome  & SE & 42 & NL & 10 & 10 & 80 &  &  & \\
\cmidrule{1-10}\pagebreak[0]
 &  & 4 & AU &  &  &  &  &  & \\
\nopagebreak
 &  & 24 & SL &  &  &  &  &  & \\
\nopagebreak
 &  & 53 & BE &  &  &  &  &  & \\
\nopagebreak
 &  & 84 & DE &  &  &  &  &  & \\
\nopagebreak
\multirow{-5}{3cm}{\raggedright\arraybackslash Hepato-pulmonary syndrome } & \multirow{-5}{*}{\raggedright\arraybackslash SE} & 104 & HR & \multirow{-5}{0.9cm}{\raggedright\arraybackslash 15} & \multirow{-5}{0.9cm}{\raggedright\arraybackslash 10} & \multirow{-5}{0.9cm}{\raggedright\arraybackslash 80} & \multirow{-5}{0.9cm}{\raggedright\arraybackslash } & \multirow{-5}{0.9cm}{\raggedright\arraybackslash } & \multirow{-5}{0.9cm}{\raggedright\arraybackslash }\\
\cmidrule{1-10}\pagebreak[0]
\rowcolor{gray!6}
 &  & 5 & AU &  &  &  &  &  & \\
\rowcolor{gray!6}
\nopagebreak
 &  & 25 & SL &  &  &  &  &  & \\
\rowcolor{gray!6}
\nopagebreak
 &  & 43 & NL &  &  &  &  &  & \\
\rowcolor{gray!6}
\nopagebreak
 &  & 54 & BE &  &  &  &  &  & \\
\rowcolor{gray!6}
\nopagebreak
 &  & 91 & DE &  &  &  &  &  & \\
\rowcolor{gray!6}
\nopagebreak
\multirow{-6}{3cm}{\raggedright\arraybackslash Porto-pulmonary hypertension } & \multirow{-6}{*}{\raggedright\arraybackslash SE} & 105 & HR & \multirow{-6}{0.9cm}{\raggedright\arraybackslash 25} & \multirow{-6}{0.9cm}{\raggedright\arraybackslash 10} & \multirow{-6}{0.9cm}{\raggedright\arraybackslash 80} & \multirow{-6}{0.9cm}{\raggedright\arraybackslash } & \multirow{-6}{0.9cm}{\raggedright\arraybackslash } & \multirow{-6}{0.9cm}{\raggedright\arraybackslash }\\
\cmidrule{1-10}\pagebreak[0]
 &  & 6 & AU &  &  &  &  &  & \\
\nopagebreak
 &  & 26 & SL &  &  &  &  &  & \\
\nopagebreak
 &  & 44 & NL &  &  &  &  &  & \\
\nopagebreak
 &  & 55 & BE &  &  &  &  &  & \\
\nopagebreak
 &  & 85 & DE &  &  &  &  &  & \\
\nopagebreak
\multirow{-6}{3cm}{\raggedright\arraybackslash Primary Hyperoxaluria Type 1 } & \multirow{-6}{*}{\raggedright\arraybackslash SE} & 106 & HR & \multirow{-6}{0.9cm}{\raggedright\arraybackslash 10} & \multirow{-6}{0.9cm}{\raggedright\arraybackslash 10} & \multirow{-6}{0.9cm}{\raggedright\arraybackslash 80} & \multirow{-6}{0.9cm}{\raggedright\arraybackslash } & \multirow{-6}{0.9cm}{\raggedright\arraybackslash 2} & \multirow{-6}{0.9cm}{\raggedright\arraybackslash }\\
\cmidrule{1-10}\pagebreak[0]
\rowcolor{gray!6}
 &  & 7 & AU &  &  &  &  &  & \\
\rowcolor{gray!6}
\nopagebreak
 &  & 27 & SL &  &  &  &  &  & \\
\rowcolor{gray!6}
\nopagebreak
 &  & 45 & NL &  &  &  &  &  & \\
\rowcolor{gray!6}
\nopagebreak
 &  & 56 & BE &  &  &  &  &  & \\
\rowcolor{gray!6}
\nopagebreak
 &  & 86 & DE &  &  &  &  &  & \\
\rowcolor{gray!6}
\nopagebreak
\multirow{-6}{3cm}{\raggedright\arraybackslash Polycystic liver disease } & \multirow{-6}{*}{\raggedright\arraybackslash SE} & 107 & HR & \multirow{-6}{0.9cm}{\raggedright\arraybackslash 10} & \multirow{-6}{0.9cm}{\raggedright\arraybackslash 10} & \multirow{-6}{0.9cm}{\raggedright\arraybackslash 80} & \multirow{-6}{0.9cm}{\raggedright\arraybackslash } & \multirow{-6}{0.9cm}{\raggedright\arraybackslash } & \multirow{-6}{0.9cm}{\raggedright\arraybackslash }\\
\cmidrule{1-10}\pagebreak[0]
Hepatocellular Carcinoma  & SE & 46 & NL & 10 & 10 & 80 &  &  & \\
\cmidrule{1-10}\pagebreak[0]
\rowcolor{gray!6}
 &  & 8 & AU &  &  &  &  &  & \\
\rowcolor{gray!6}
\nopagebreak
 &  & 28 & SL &  &  &  &  &  & \\
\rowcolor{gray!6}
\nopagebreak
 &  & 57 & BE &  &  &  &  &  & \\
\rowcolor{gray!6}
\nopagebreak
 &  & 87 & DE &  &  &  &  &  & \\
\rowcolor{gray!6}
\nopagebreak
\multirow{-5}{3cm}{\raggedright\arraybackslash Hepatocellular Carcinoma } & \multirow{-5}{*}{\raggedright\arraybackslash SE} & 108 & HR & \multirow{-5}{0.9cm}{\raggedright\arraybackslash 15} & \multirow{-5}{0.9cm}{\raggedright\arraybackslash 10} & \multirow{-5}{0.9cm}{\raggedright\arraybackslash 80} & \multirow{-5}{0.9cm}{\raggedright\arraybackslash } & \multirow{-5}{0.9cm}{\raggedright\arraybackslash } & \multirow{-5}{0.9cm}{\raggedright\arraybackslash }\\
\cmidrule{1-10}\pagebreak[0]
 &  & 9 & AU &  &  &  &  &  & \\
\nopagebreak
 &  & 29 & SL &  &  &  &  &  & \\
\nopagebreak
 &  & 47 & NL &  &  &  &  &  & \\
\nopagebreak
 &  & 58 & BE &  &  &  &  &  & \\
\nopagebreak
 &  & 88 & DE &  &  &  &  &  & \\
\nopagebreak
 &  & 109 & HR &  &  &  &  &  & \\
\nopagebreak
\multirow{-7}{3cm}{\raggedright\arraybackslash Non-metastatic hepatoblastoma} & \multirow{-7}{*}{\raggedright\arraybackslash SE} & 151 & HU & \multirow{-7}{0.9cm}{\raggedright\arraybackslash 50} & \multirow{-7}{0.9cm}{\raggedright\arraybackslash 0} & \multirow{-7}{0.9cm}{\raggedright\arraybackslash 80} & \multirow{-7}{0.9cm}{\raggedright\arraybackslash } & \multirow{-7}{0.9cm}{\raggedright\arraybackslash } & \multirow{-7}{0.9cm}{\raggedright\arraybackslash }\\
\cmidrule{1-10}\pagebreak[0]
\rowcolor{gray!6}
 &  & 10 & AU &  &  &  &  &  & \\
\rowcolor{gray!6}
\nopagebreak
 &  & 30 & SL &  &  &  &  &  & \\
\rowcolor{gray!6}
\nopagebreak
 &  & 48 & NL &  &  &  &  &  & \\
\rowcolor{gray!6}
\nopagebreak
 &  & 59 & BE &  &  &  &  &  & \\
\rowcolor{gray!6}
\nopagebreak
 &  & 89 & DE &  &  &  &  &  & \\
\rowcolor{gray!6}
\nopagebreak
 &  & 110 & HR &  &  &  &  &  & \\
\rowcolor{gray!6}
\nopagebreak
\multirow{-7}{3cm}{\raggedright\arraybackslash Urea cycle disorder/organic acidemia} & \multirow{-7}{*}{\raggedright\arraybackslash SE} & 152 & HU & \multirow{-7}{0.9cm}{\raggedright\arraybackslash 50} & \multirow{-7}{0.9cm}{\raggedright\arraybackslash 0} & \multirow{-7}{0.9cm}{\raggedright\arraybackslash 80} & \multirow{-7}{0.9cm}{\raggedright\arraybackslash } & \multirow{-7}{0.9cm}{\raggedright\arraybackslash } & \multirow{-7}{0.9cm}{\raggedright\arraybackslash }\\
\cmidrule{1-10}\pagebreak[0]
 &  & 11 & AU &  &  &  &  &  & \\
\nopagebreak
 &  & 31 & SL &  &  &  &  &  & \\
\nopagebreak
 &  & 49 & NL &  &  &  &  &  & \\
\nopagebreak
 &  & 60 & BE &  &  &  &  &  & \\
\nopagebreak
 &  & 90 & DE &  &  &  &  &  & \\
\nopagebreak
\multirow{-6}{3cm}{\raggedright\arraybackslash Persistent hepatic dysfunction } & \multirow{-6}{*}{\raggedright\arraybackslash SE} & 111 & HR & \multirow{-6}{0.9cm}{\raggedright\arraybackslash } & \multirow{-6}{0.9cm}{\raggedright\arraybackslash } & \multirow{-6}{0.9cm}{\raggedright\arraybackslash } & \multirow{-6}{0.9cm}{\raggedright\arraybackslash 20} & \multirow{-6}{0.9cm}{\raggedright\arraybackslash } & \multirow{-6}{0.9cm}{\raggedright\arraybackslash }\\
\cmidrule{1-10}\pagebreak[0]
\rowcolor{gray!6}
 &  & 114 & NL &  &  &  &  &  & \\
\rowcolor{gray!6}
\nopagebreak
 &  & 115 & AU &  &  &  &  &  & \\
\rowcolor{gray!6}
\nopagebreak
 &  & 116 & SL &  &  &  &  &  & \\
\rowcolor{gray!6}
\nopagebreak
 &  & 117 & BE &  &  &  &  &  & \\
\rowcolor{gray!6}
\nopagebreak
 &  & 118 & DE &  &  &  &  &  & \\
\rowcolor{gray!6}
\nopagebreak
\multirow{-6}{3cm}{\raggedright\arraybackslash Hereditary hemorrhagic teleangiectasia } & \multirow{-6}{*}{\raggedright\arraybackslash SE} & 119 & HR & \multirow{-6}{0.9cm}{\raggedright\arraybackslash 15} & \multirow{-6}{0.9cm}{\raggedright\arraybackslash 10} & \multirow{-6}{0.9cm}{\raggedright\arraybackslash 80} & \multirow{-6}{0.9cm}{\raggedright\arraybackslash } & \multirow{-6}{0.9cm}{\raggedright\arraybackslash } & \multirow{-6}{0.9cm}{\raggedright\arraybackslash }\\
\cmidrule{1-10}\pagebreak[0]
 &  & 121 & AU &  &  &  &  &  & \\
\nopagebreak
 &  & 122 & SL &  &  &  &  &  & \\
\nopagebreak
 &  & 123 & BE &  &  &  &  &  & \\
\nopagebreak
 &  & 124 & DE &  &  &  &  &  & \\
\nopagebreak
 &  & 125 & HR &  &  &  &  &  & \\
\nopagebreak
\multirow{-6}{3cm}{\raggedright\arraybackslash Hepatic hemangioendothelioma} & \multirow{-6}{*}{\raggedright\arraybackslash SE} & 126 & NL & \multirow{-6}{0.9cm}{\raggedright\arraybackslash 15} & \multirow{-6}{0.9cm}{\raggedright\arraybackslash 10} & \multirow{-6}{0.9cm}{\raggedright\arraybackslash 80} & \multirow{-6}{0.9cm}{\raggedright\arraybackslash } & \multirow{-6}{0.9cm}{\raggedright\arraybackslash } & \multirow{-6}{0.9cm}{\raggedright\arraybackslash }\\
\cmidrule{1-10}\pagebreak[0]
\rowcolor{gray!6}
 &  & 131 & SL &  &  &  &  &  & \\
\rowcolor{gray!6}
\nopagebreak
\multirow{-2}{3cm}{\raggedright\arraybackslash Biliary sepsis} & \multirow{-2}{*}{\raggedright\arraybackslash SE} & 133 & HR & \multirow{-2}{0.9cm}{\raggedright\arraybackslash } & \multirow{-2}{0.9cm}{\raggedright\arraybackslash } & \multirow{-2}{0.9cm}{\raggedright\arraybackslash } & \multirow{-2}{0.9cm}{\raggedright\arraybackslash 20} & \multirow{-2}{0.9cm}{\raggedright\arraybackslash } & \multirow{-2}{0.9cm}{\raggedright\arraybackslash }\\
\cmidrule{1-10}\pagebreak[0]
Biliary sepsis & SE & 134 & NL &  &  &  & 30 &  & \\
\cmidrule{1-10}\pagebreak[0]
\rowcolor{gray!6}
Biliary sepsis/ Secondary sclerosing cholangitis  & SE & 132 & DE &  &  &  & 30 &  & \\
\cmidrule{1-10}\pagebreak[0]
Primary sclerosing cholangitis  & SE & 142 & DE & 15 & 10 & 80 &  &  & \\
\cmidrule{1-10}\pagebreak[0]
\rowcolor{gray!6}
 &  & 141 & SL &  &  &  &  &  & \\
\rowcolor{gray!6}
\nopagebreak
\multirow{-2}{3cm}{\raggedright\arraybackslash Primary sclerosing cholangitis } & \multirow{-2}{*}{\raggedright\arraybackslash SE} & 143 & HR & \multirow{-2}{0.9cm}{\raggedright\arraybackslash } & \multirow{-2}{0.9cm}{\raggedright\arraybackslash } & \multirow{-2}{0.9cm}{\raggedright\arraybackslash } & \multirow{-2}{0.9cm}{\raggedright\arraybackslash 20} & \multirow{-2}{0.9cm}{\raggedright\arraybackslash } & \multirow{-2}{0.9cm}{\raggedright\arraybackslash }\\
\cmidrule{1-10}\pagebreak[0]
Primary sclerosing cholangitis  & SE & 144 & NL &  &  &  & 30 &  & \\
\cmidrule{1-10}\pagebreak[0]
\rowcolor{gray!6}
 &  & 145 & BE &  &  &  &  &  & \\
\rowcolor{gray!6}
\nopagebreak
 &  & 146 & NL &  &  &  &  &  & \\
\rowcolor{gray!6}
\nopagebreak
 &  & 147 & HR &  &  &  &  &  & \\
\rowcolor{gray!6}
\nopagebreak
 &  & 148 & SL &  &  &  &  &  & \\
\rowcolor{gray!6}
\nopagebreak
 &  & 149 & AU &  &  &  &  &  & \\
\rowcolor{gray!6}
\nopagebreak
\multirow{-6}{3cm}{\raggedright\arraybackslash Biliary atresia} & \multirow{-6}{*}{\raggedright\arraybackslash SE} & 150 & DE & \multirow{-6}{0.9cm}{\raggedright\arraybackslash 60} & \multirow{-6}{0.9cm}{\raggedright\arraybackslash 15} & \multirow{-6}{0.9cm}{\raggedright\arraybackslash 80} & \multirow{-6}{0.9cm}{\raggedright\arraybackslash } & \multirow{-6}{0.9cm}{\raggedright\arraybackslash 1} & \multirow{-6}{0.9cm}{\raggedright\arraybackslash }\\
\cmidrule{1-10}\pagebreak[0]
Neuro Endocrine Tumor & SE & 160 & DE & 15 & 10 & 80 &  &  & \\
\cmidrule{1-10}\pagebreak[0]
\rowcolor{gray!6}
 &  & 172 & BE &  &  &  &  &  & \\
\rowcolor{gray!6}
\nopagebreak
\multirow{-2}{3cm}{\raggedright\arraybackslash Hepatic artery thrombosis} & \multirow{-2}{*}{\raggedright\arraybackslash SE} & 174 & NL & \multirow{-2}{0.9cm}{\raggedright\arraybackslash 50} & \multirow{-2}{0.9cm}{\raggedright\arraybackslash 0} & \multirow{-2}{0.9cm}{\raggedright\arraybackslash 80} & \multirow{-2}{0.9cm}{\raggedright\arraybackslash } & \multirow{-2}{0.9cm}{\raggedright\arraybackslash } & \multirow{-2}{0.9cm}{\raggedright\arraybackslash }\\
\cmidrule{1-10}\pagebreak[0]
 &  & 177 & DE &  &  &  &  &  & \\
\nopagebreak
 &  & 171 & AU &  &  &  &  &  & \\
\nopagebreak
 &  & 173 & HU &  &  &  &  &  & \\
\nopagebreak
 &  & 175 & HR &  &  &  &  &  & \\
\nopagebreak
\multirow{-5}{3cm}{\raggedright\arraybackslash Hepatic artery thrombosis} & \multirow{-5}{*}{\raggedright\arraybackslash SE} & 176 & SL & \multirow{-5}{0.9cm}{\raggedright\arraybackslash 100} & \multirow{-5}{0.9cm}{\raggedright\arraybackslash 0} & \multirow{-5}{0.9cm}{\raggedright\arraybackslash 80} & \multirow{-5}{0.9cm}{\raggedright\arraybackslash } & \multirow{-5}{0.9cm}{\raggedright\arraybackslash } & \multirow{-5}{0.9cm}{\raggedright\arraybackslash }\\
\cmidrule{1-10}\pagebreak[0]
\rowcolor{gray!6}
 &  & 701 & DE &  &  &  &  &  & \\
\rowcolor{gray!6}
\nopagebreak
 &  & 702 & AU &  &  &  &  &  & \\
\rowcolor{gray!6}
\nopagebreak
 &  & 703 & BE &  &  &  &  &  & \\
\rowcolor{gray!6}
\nopagebreak
 &  & 704 & HR &  &  &  &  &  & \\
\rowcolor{gray!6}
\nopagebreak
 &  & 705 & HU &  &  &  &  &  & \\
\rowcolor{gray!6}
\nopagebreak
 &  & 706 & NL &  &  &  &  &  & \\
\rowcolor{gray!6}
\nopagebreak
\multirow{-7}{3cm}{\raggedright\arraybackslash PED (12-16 years)} & \multirow{-7}{*}{\raggedright\arraybackslash PED} & 707 & SL & \multirow{-7}{0.9cm}{\raggedright\arraybackslash 15} & \multirow{-7}{0.9cm}{\raggedright\arraybackslash 10} & \multirow{-7}{0.9cm}{\raggedright\arraybackslash 90} & \multirow{-7}{0.9cm}{\raggedright\arraybackslash } & \multirow{-7}{0.9cm}{\raggedright\arraybackslash 16} & \multirow{-7}{0.9cm}{\raggedright\arraybackslash }\\
\cmidrule{1-10}\pagebreak[0]
 &  & 801 & DE &  &  &  &  &  & 701\\
\nopagebreak
 &  & 802 & AU &  &  &  &  &  & 702\\
\nopagebreak
 &  & 803 & BE &  &  &  &  &  & 703\\
\nopagebreak
 &  & 804 & HR &  &  &  &  &  & 704\\
\nopagebreak
 &  & 805 & SL &  &  &  &  &  & 705\\
\nopagebreak
 &  & 806 & HU &  &  &  &  &  & 706\\
\nopagebreak
\multirow{-7}{3cm}{\raggedright\arraybackslash PED ($<$12 years)} & \multirow{-7}{*}{\raggedright\arraybackslash PED} & 807 & NL & \multirow{-7}{0.9cm}{\raggedright\arraybackslash 35} & \multirow{-7}{0.9cm}{\raggedright\arraybackslash 15} & \multirow{-7}{0.9cm}{\raggedright\arraybackslash 90} & \multirow{-7}{0.9cm}{\raggedright\arraybackslash } & \multirow{-7}{0.9cm}{\raggedright\arraybackslash 12} & 707\\
\cmidrule{1-10}\pagebreak[0]
\rowcolor{gray!6}
 &  & 902 & AU &  &  &  &  &  & \\
\rowcolor{gray!6}
\nopagebreak
 &  & 904 & HR &  &  &  &  &  & \\
\rowcolor{gray!6}
\nopagebreak
 &  & 905 & HU &  &  &  &  &  & \\
\rowcolor{gray!6}
\nopagebreak
 &  & 906 & NL &  &  &  &  &  & \\
\rowcolor{gray!6}
\nopagebreak
\multirow{-5}{3cm}{\raggedright\arraybackslash NSE} & \multirow{-5}{*}{\raggedright\arraybackslash NSE} & 907 & SL & \multirow{-5}{0.9cm}{\raggedright\arraybackslash 10} & \multirow{-5}{0.9cm}{\raggedright\arraybackslash 10} & \multirow{-5}{0.9cm}{\raggedright\arraybackslash 80} & \multirow{-5}{0.9cm}{\raggedright\arraybackslash } & \multirow{-5}{0.9cm}{\raggedright\arraybackslash } & \multirow{-5}{0.9cm}{\raggedright\arraybackslash }\\
\cmidrule{1-10}\pagebreak[0]
NSE & NSE & 903 & BE & 10 & 10 & 90 &  &  & \\
\cmidrule{1-10}\pagebreak[0]
\rowcolor{gray!6}
NSE & NSE & 901 & DE & 15 & 10 & 80 &  &  & 
\end{longtable}

\endgroup

\FloatBarrier
\vspace{.1cm}
\newpage

\section{Details to the future status update imputation procedure}\label{app:imputation}

\paragraph*{\texorpdfstring{Overview of the \emph{counterfactual} future status imputation procedure}{Overview of the counterfactual future status imputation procedure}}\label{overview-of-the-counterfactual-future-status-imputation-procedure}
\addcontentsline{toc}{paragraph}{Overview of the \emph{counterfactual} future status imputation procedure}

The ELAS Simulator requires a complete stream of status updates for all
patient registrations, i.e.~all registrations must end with a removal
(R) status or a waiting list death (D) status. Input streams for
registrations which end with a transplantation (FU) thus have to be
completed with synthetic status updates. For LSAM, this problem was
addressed by matching transplanted patients to non-transplanted patients
on expected mortality, and copying over future status updates from a
matched candidate. Potential limitations of this approach are that
patients were not matched based on other characteristics such as disease
groups , and that transplantation is a dependent censoring mechanism
which results in biased predictions of expected mortality.

To address these limitations, we implemented a procedure for the ELAS
simulator which matches transplanted candidates to candidates still at
risk based on their expected \emph{counterfactual} mortality risk and on
other characteristics (such as HU status, age, and type of liver
disease). In predicting this counterfactual mortality, we correct for
dependent censoring by transplantation with inverse probability
censoring weighting by modifying a procedure proposed by Tayob and
Murray to impute future survival times from pre-determined landmark
times for lung waiting list transplantation candidates.

Specifically, Tayob and Murray are interested in modelling the 12-month
restricted survival time \(T^* = \min(T, \tau)\) of transplant candidates,
where \(T\) is a random variable for the remaining survival time and
\(\tau\) is the time horizon of interest (12 months in their case). Tayob
and Murray work with regularly spaced landmark times \(j\), and define
\(T^*_{ij}\) as the \(\tau\)-restricted remaining survival time of subject
\(i\) from landmark time \(j\). With these, Tayob and Murray aim to model
the expected mean survival time directly, i.e.~model
\[\mathbb{E}[\log(T^*) | Z] = \beta^\intercal Z.\] A challenge for
direct estimation of \(\beta\) is that \(T^*_{ij}\) is unobserved for most
candidates due to censoring, transplantation represents an informative
censoring mechanism, and that the \(T^*_{ij}\) exhibit correlations over
the landmark times \(j\).

Tayob and Murray propose to multiply impute \(T_{ij}^*\) for all \(i\) and
\(j\)s, and to estimate \(\beta\) with Generalized Estimating Equations
(GEE) to account for the correlation between \(T_{ij}^*\). In summary,
Tayob and Murray's imputation procedure consists of:

\begin{enumerate}
\def\labelenumi{\arabic{enumi}.}
\item
  Estimate the counterfactual waiting list survival function
  \(S^{\text{IPCW}}_T(t)\), with correction for informative censoring by
  transplantation with inverse probability censoring weighting (IPCW).
  With this curve, they can construct \(j\)-specific pseudo-observations
  \(PO_{ij}\) for \(\log(T^*_{ij})\).
\item
  Estimate \(\hat{\beta}^{\text{PO}}\) on pairs \((PO_{ij}, Z_{ij})\) with
  Generalized Estimating Equations (GEE) with an unstructured working
  correlation matrix. This correlation matrix permits arbitrary
  correlation between the \(T_{ij}\)s over \(j\).
\item
  For each patient \(i\) with unobserved \(T^*_{ij}\) construct a risk set
  \(R_i\)

  \begin{itemize}
  \item
    of patients who remain at risk after patient \(i\) is censored,
    i.e.~with \(T_{kj} > C_{ij}\), \(k\neq i\), where \(C_{ij}\) is
    patient \(i\)'s censoring time measured from landmark time \(j\),
  \item
    who are similar in terms of predicted expected log survival,
    i.e.~require
    \[|\hat{\beta}^{\text{PO}}\ ^\intercal Z_k(C_{i}) - \hat{\beta}^{\text{PO}}\ ^\intercal Z_i(C_{i})| < \epsilon\]
    for some \(\epsilon\), where \(Z_k(C_{i})\) are covariates of
    patient \(k\) at \(i\)'s censoring time \(C_i\),
  \item
    Optionally, further restrict matches based on \(Z_i(C_{i})\), e.g.
    require matching disease groups.
  \end{itemize}
\item
  Within risk set \(R_i\), re-estimate counterfactual survival function
  \(S_T^{\text{IPCW}}(t|R_i)\). Use inverse transform sampling from
  \(\hat{S}^{\text{IPCW}}_T(t|R_i)\) to sample a valid \(T^*_{ij}.\)
\end{enumerate}

Our aim is to use Tayob and Murray's procedure to match transplant
recipients to comparable, not-yet-transplanted waiting list candidates.
This requires two main modifications to the imputation procedure:

\begin{itemize}
\item
  We have to match transplanted patients at any continuous time \(t\) to
  comparable, at-risk waiting list candidates (and not only at
  pre-determined landmark times \(j\)). For this, we define \(T^*_{it}\)
  to be the \(\tau\)-restricted remaining survival time from time \(t\)
  forward, with \(t\) the number of days elapsed since listing. For each
  observed status update we construct a \(t\)-specific
  pseudo-observation \(PO_{it}\), with \(t\) the time-after-listing at
  which the status was reported.
\item
  We cannot estimate \({\beta}^{\text{PO}}\) with GEE with an
  unstructured correlation matrix, as the \(PO_{it}\) is indexed by
  continuous time \(t\). Instead, we propose to estimate \(\beta^{PO}\)
  with Quasi-Least Squares (QLS) with a Markov correlation structure .
  This structure assumes that the correlation between \(PO_{it}\) decays
  with spacing in \(t\).
\end{itemize}

Construction of the risk set \(R_i\) for patients with unknown \(T^*_{it}\)
is then similar; we consider patients \(k\) with \(T^*_{kt} > C^*_{it}\),
similar \(\hat{\beta}^{\text{PO}}\ ^\intercal Z_\cdot(C_{ik})\), and
require matches on characteristics \(Z_{ik}(C_i)\). For each risk set, we
can estimate \(S^{\text{IPCW}}_T(t | R_i)\). Inverse transform sampling
from \(S^{\text{IPCW}}_T(t)\) is then used to match candidate \(i\) to a
specific candidate \(k \in R_i\). We can then impute patient \(i\)'s future
status updates by copying over status updates from patient \(k\). This
procedure is repeated, until all candidates have a set of status updates
ending with a waiting list removal (R) or death (D).

This proposed imputation procedure was ran separately for HU and T
candidates. Repeatedly running the procedure results in different input
status files for the ELAS simulator. In the remainder of this appendix,
we describe elements of the proposed imputation procedure in detail, and
describe which variables were adjusted for in each step of the
imputation procedure.

\subsection{IPCW survival curve estimate \& construction of pseudo-outcomes}\label{ipcw-survival-curve-estimate-construction-of-pseudo-outcomes}

Here, we discuss the procedure used to estimate counterfactual survival
curves, and how to construct pseudo-outcomes for the log expected
survival times.

\vspace*{.5cm}

\subsubsection*{Propensity score model}

\hfill\break
We are interested estimation of counterfactual waiting list survival,
i.e.~the probability that a patient is not yet delisted / has not yet
died on the waiting list if transplantation were not available.
Problematic is that in reality we are less likely to observe waiting
list removals / deaths for patients who deteriorate, as MELD-based liver
allocation offers livers first to the most sick patients, making
transplantation an informative censoring mechanism. To correct for
informative censoring by transplantation, we use a Cox model to predict
the probability that a patient is censored over time, and estimate the
counterfactual waiting list survival curve weighing observations by the
inverse probability of being transplanted.

Adjustment variables included in the Cox model for prediction of the
transplant probability are recipient sex, recipient blood group, spline
terms of recipient weight and age, recipient disease group, percentage
of time NT (total/too bad/too good), the national match MELD, whether
the patient is on dialysis, whether the patient has a downmarked MELD
and whether the patient has an exception (Y/N).

\begin{figure}[h]

{\centering \includegraphics[width=0.9\linewidth]{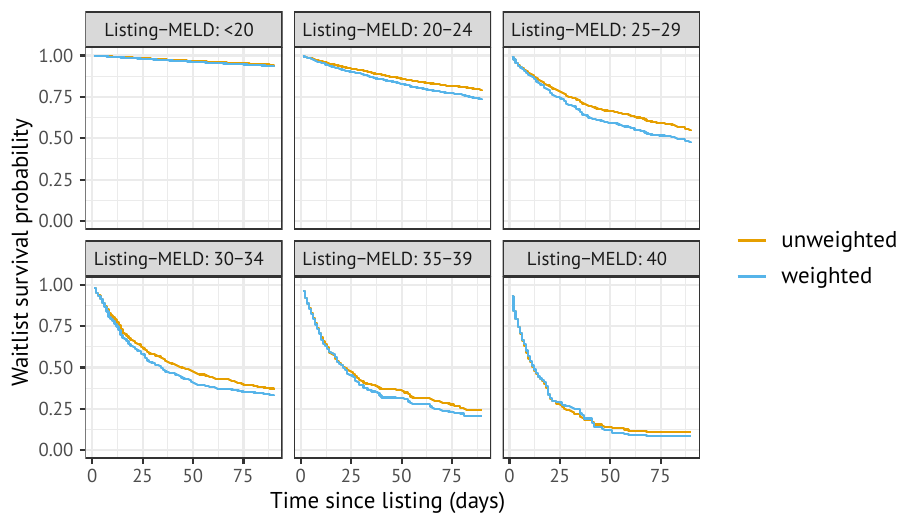} 

}

\caption{Estimated survival probabilities estimated in the cohort, with (blue) and without (orange) Inverse Probability Censoring Weighting to correct for informative censoring by transplantation.}\label{fig:chappsfig2}
\end{figure}

Figure \ref{fig:chappsfig2} shows estimated survival functions
without (orange) and with (blue) correction for dependent censoring for
non-HU patients, stratified by the laboratory MELD score at listing. As
expected, estimated 90-day survival probabilities decrease due to
inverse probability weighting, with 90-day waiting list survival
estimated with IPCW up to 7.4\% lower for MELD 25--29 than 90-day waiting
list survival estimated without IPCW.

\vspace*{.5cm}
\noindent
\textbf{Construction of pseudo-outcomes}\\
As in , we are interested in directly modeling residual remaining
survival time \(T^{*}\), i.e.~modelling
\[\mathbb{E}[\log(T^*)|Z] = \beta^\intercal Z.\] For this, we'd ideally
have access to observed times-to-event \(T^*_{it}\), i.e.~the
time-to-event for patient \(i\) after time \(t\) (where \(t\) is a time at
which the patient reported a status update). However, censoring and
transplantation within \(\tau\) time units of \(t\) prevent us from
observing \(T^*_{it}\). Tayob and Murray described how pseudo-observations
for \(\log(T_{ij}*)\) can be constructed, based only on the estimated
counterfactual survival curve \(\hat{S}^{IPCW}(t)\). We used this
procedure to construct pseudo-observations for \(\log(T_{it}*)\). Armed
with pairs \((PO_{it}, Z_i)\), we can in principle estimate \(\beta^{PO}\).

\subsection{Modelling the mean restricted survival time with Quasi-Least Squares}\label{modelling-the-mean-restricted-survival-time-with-quasi-least-squares}

With pairs \((PO_{it}, Z_i)\) we can model the expected log remaining
survival time as \[\mathbb{E}[\log(T^*) | Z] = \beta^\intercal Z
    \label{eqn:model_tayob}\] However, in estimation of \(\beta\) requires
that we deal with correlations in \(PO_{it}\) over \(t\). Tayob and Murray
faced a similar issue, where there is correlation between \(T^*_{ij}\) for
the different landmark times \(j\), and addressed this by estimating
\(\beta\) with Generalized Estimating Equations (GEE) with an unstructured
correlation matrix correlation over the landmark times \(j\) (which
requires \(j(j+1)/2\) parameters). Unfortunately, this strategy is
infeasible to us as potential outcomes \(PO_{it}\) are indexed by
continuous time \(t\), i.e.~all observed censoring times. Instead, we
assume a Markov correlation structure for the potential outcomes, i.e.
we assume that the correlation between measurements \(PO_{is}\) and
\(PO_{it}\) decays with their separation in time:
\[\texttt{Corr}(PO_{is}, PO_{it}) = \alpha^{|s-t|}.\] Parameters
\(\alpha\) and \(\beta\) to this model may be estimated with the \texttt{qlspack} R
package, with Quasi-Least Squares.

We use different model specifications for HU and non-HU patients for
equation, with adjustment for natural spline terms
for continuous variables. For HU patients, adjustment variables include
recipient age at registration, the laboratory MELD score, whether the
patient is on biweekly dialysis, recipient sex, and disease group. For
elective patients, we adjust for age at registration, recipient weight,
MELD components (serum creatinine, bilirubin, INR, biweekly dialysis),
recipient sex, disease group, cirrhosis aetiology, type of exception
score, whether it is a retransplant candidate, and whether the patient
has failed to recertify their MELD score.

\subsection{Risk set construction}\label{risk-set-construction}

As in Tayob and Murray, a minimal requirement to match transplanted
candidates to not-yet-transplanted candidates is:
\(|\hat{\beta}^{\text{PO}}\ ^\intercal Z_k(C_{i}) - \hat{\beta}^{\text{PO}}\ ^\intercal Z_i(C_{i})| < 0.50\),
i.e.~candidates have similar expected log 90-day-truncated survival. For
both HU patients and non-HU patients, we require to always match on
pediatric status. For other discrete variables and continuous variables
we use an adaptive matching procedure in which we strive towards 35
candidates in the risk set for HU patients, and 50 candidates for non-HU
patients.

Specifically, the discrete variables used for matching are

\begin{enumerate}
\def\labelenumi{\arabic{enumi}.}
\item
  whether the patient is a retransplant candidate
\item
  current urgency code (non-transplantable)
\item
  (N)SE group
\item
  disease group
\item
  urgency reason (NT too good / NT other / NT too bad)
\item
  biweekly dialysis (twice in week preceding MELD measurement)
\item
  recipient country.
\end{enumerate}

Continuous match variables used are the laboratory MELD score, age at
registration, (N)SE MELD score (for elective patients only), where we
restrict absolute differences in continuous variables to pre-determined
caliper widths (lab-MELD: 5, age: 15 years, (N)SE-MELD: 5). In case
matching according to all criteria fails to result in a risk set of
sufficient size, we drop a discrete match criterion (from 7 to 1 in the
list above). In case dropping all discrete match criteria does not
result in adequately sized risk set, we increase caliper widths for
continuous variables. In total, about 50\% of transplant recipients can
be matched to a risk set on all characteristics, and 80\% is matchable on
the first 4 discrete variables (with the most restrictive caliper
widths).

\vspace*{.5cm}
\noindent
\textbf{Example of a risk set}\\
Table \ref{tab:apptab1} shows an example of a constructed
risk set for a patient who was transplanted, and for whom we thus need
to impute future status updates. The first row of Table
\ref{tab:apptab1} shows that the transplanted patient
is listed in 2014 in Germany at an age of 64 for liver cirrhosis. The
patient reported a 33 lab-MELD score 36 days after registration, and was
transplanted 6 days later. Based on our model, the expected log residual
survival time for this patient is approximately 3.51, which corresponds
roughly to 34 days.

Remaining rows of Table \ref{tab:apptab1} show 10 (out of 50) waiting list
candidates in patient \(i\)'s risk set \(R_i\). These patients remain at
risk 36 days after waiting list registration (\(\min(C_{k}, T_k) > C_i\))
and are similar in terms of predicted expected log survival. Turning to
other characteristics, we see that the matching procedure did not match
on listing country and receival of biweekly dialysis. The risk set is
comparable in terms of continuous variables (lab-MELDs ranging from 28
to 38, ages from 53 to 69).

\begin{table}[h]
\centering
\caption{\label{tab:apptab1}Example of the matched risk set for a selected liver transplantation recipient.}
\centering
\resizebox{\ifdim\width>\linewidth\linewidth\else\width\fi}{!}{
\fontsize{10}{12}\selectfont
\begin{tabular}[t]{llllllllllllll}
\toprule
year & status time $t$ & $C_k$ & $T_k$ & $\exp(\hat{\log(T^*)})$ & lab-MELD & age & ped. & reTX & urg & (N)SE & diag. & dial. & country\\
\midrule
\addlinespace[0.3em]
\multicolumn{14}{l}{\textbf{To be imputed candidate}}\\
\hspace{1em}2014 & 36.0 & 41.9 & - & 33.7 & 33 & 64 & 0 & 0 & T & none & Cirrh. & 1 & DE\\
\addlinespace[0.3em]
\multicolumn{14}{l}{\textbf{Risk set}}\\
\hspace{1em}2012 & 40.1 & 46 & - & 32.8 & 34 & 65 & 0 & 0 & T & none & Cirrh. & 0 & DE\\
\hspace{1em}2015 & 41.1 & 55.8 & - & 34.9 & 30 & 63 & 0 & 0 & T & none & Cirrh. & 1 & DE\\
\hspace{1em}2013 & 40.1 & 47.9 & - & 32.4 & 38 & 56 & 0 & 0 & T & none & Cirrh. & 1 & DE\\
\hspace{1em}2012 & 36.4 & 42.1 & - & 35.3 & 35 & 68 & 0 & 0 & T & none & Cirrh. & 1 & BE\\
\hspace{1em}2010 & 40.2 & 58.9 & - & 35.5 & 28 & 68 & 0 & 0 & T & none & Cirrh. & 0 & DE\\
\hspace{1em}2016 & 37.7 & 86.3 & - & 35.6 & 34 & 53 & 0 & 0 & T & none & Cirrh. & 1 & DE\\
\hspace{1em}2009 & 41.8 & - & 84.8 & 36.3 & 33 & 53 & 0 & 0 & T & none & Cirrh. & 0 & DE\\
\hspace{1em}2008 & 35.1 & 42.1 & - & 29.9 & 30 & 69 & 0 & 0 & T & none & Cirrh. & 0 & DE\\
\hspace{1em}2012 & 37.3 & 44.9 & - & 38.4 & 37 & 64 & 0 & 0 & T & none & Cirrh. & 0 & BE\\
\hspace{1em}2010 & 7.4 & - & 946.9 & 29.4 & 29 & 63 & 0 & 0 & T & none & Cirrh. & 0 & AU\\
\bottomrule
\end{tabular}}
\end{table}

\begin{figure}[h]

{\centering \includegraphics[width=0.5\linewidth]{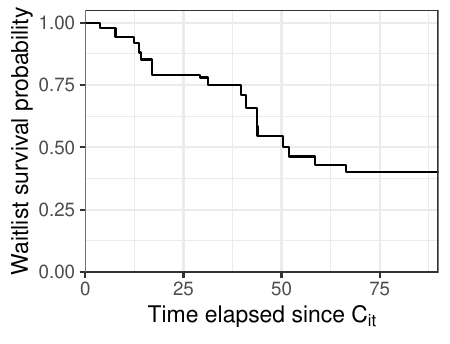} 

}

\caption{Conditional survival function, estimated with inverse probability censoring weighting for the risk set $R_i$. A subset of 10 out of 50 patients in $R_i$ were shown in Table}\label{fig:chappsfig3}
\end{figure}

With risk set \(R_i\) and constructed IPCW weights, we can obtain a
personalized estimate of the conditional probability of remaining on the
waiting list \(t\) time units after \(i\)'s censoring time (i.e.
\(\hat{S}^{\text{IPCW}}_T(t|R_i, T > C_i)\)). Estimation of this survival
function for the risk set of \ref{tab:apptab1} yields the survival function shown
in Figure \ref{fig:chappsfig3}. This shows that our subject would have an
approximately 60\% probability of waiting list death/removal in the next
90 days, if not transplanted.

\FloatBarrier

\subsection{\texorpdfstring{Matching the patient to a particular patient in \(R_i\)}{Matching the patient to a particular patient in R\_i}}\label{matching-the-patient-to-a-particular-patient-in-r_i}

We match the censored patient \(i\) to a patient who remains at risk in
the risk set \(R_i\) with inverse transform sampling. This means that we
(i) draw a random number \(u\) from the uniform(0,1) distribution, and
(ii) find the smallest matching \(t\) such that
\(\hat{S}^{\text{IPCW}}_T(t|R_i, T > C_i) \leq u\)). If such a \(t\) exists,
it corresponds to an event time for a patient \(k \in R_i\). We can impute
future status updates for patient \(i\) by copying over the future status
updates from patient \(k\).

In case such a \(t\) does not exist, the patient remains alive at least
\(\tau\) time units after the status update. We therefore need to match to
a patient from the set of patients who remain alive \(\tau\) days after
\(C_i\), i.e.~\(\{k \in R_i : T_{k} > C_i + \tau\}\). For this, we sampled
patients with probability proportional to their inverse probability
censoring weight, i.e.~proportional to
\(\mathbb{P}[T_{k\tau} \leq C_{it} + 90 | T_{k\tau} > C_{it}].\)

\subsection{Heterogeneity in waiting list survival per risk set}\label{heterogeneity-in-waiting-list-survival-per-risk-set}

\begin{figure}[h]

{\centering \includegraphics[width=0.9\linewidth]{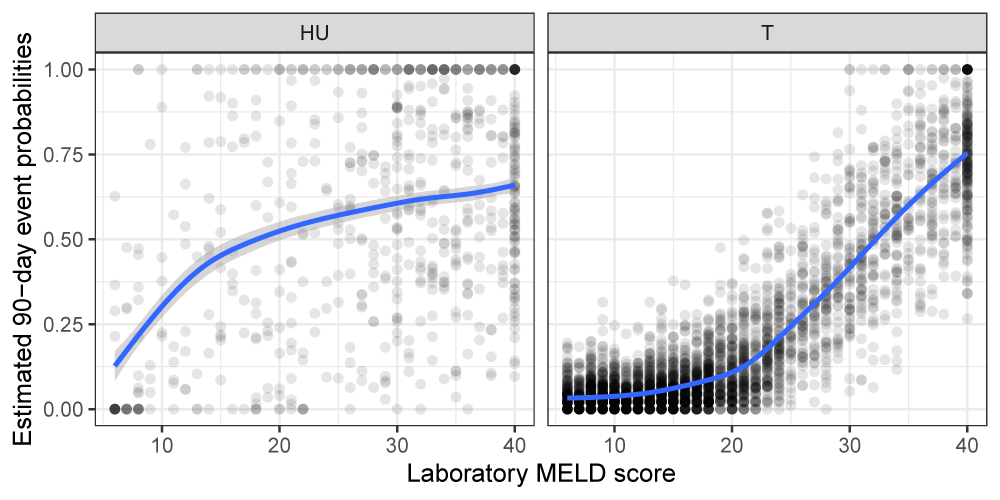} 

}

\caption{Estimates of 90-day residual survival ($\hat{S}^{\text{IPCW}}_T(90)$) per constructed risk set.}\label{fig:chappsfig4}
\end{figure}

Figure \ref{fig:chappsfig4} shows distributions of the
estimated 90-day waiting list death/removal probabilities by laboratory
MELD score, separately for HU and non-HU patients. The blue line is a
smoothed estimate of the relation between laboratory MELD score and
average 90-day event rates. The spread around the blue lines suggests
that there is substantial heterogeneity in event rates not captured by
MELD. This is especially true for HU patients (for whom MELD score is
also not used in allocation). An interesting observation is that waiting
list death/removal rates for the highest MELD scores (35--40) are higher
than HU patients, even though HU patients receive absolute priority in
allocation.

\FloatBarrier
\vspace{.1cm}
\newpage

\section{Deviation from standard allocation in the ELAS simulator's graft offering module}\label{app:devregacc}

To avoid loss of
transplantable grafts, ET can initiate the extended allocation procedure
or competitive rescue allocation . In total, such deviation from
standard allocation accounts for 25\% of placements. Without
approximating such deviation from standard allocation, too many grafts
are placed nationally in Germany (instead of regionally{[}\^{}6{]}), and too
many grafts are placed nationally in Belgium (instead of locally{[}\^{}7{]}).
Moreover, without deviation from standard allocation too few organs are
placed at lower ranks, because many candidates specify that they do not
want to receive a graft through extended allocation profiles. This
motivated us to approximate deviation from standard allocation in the
graft offering module.\\
We approximate deviation from standard allocation by

\begin{enumerate}
\def\labelenumi{\arabic{enumi}.}
\item
  simulating the maximum number of offers made until rescue allocation
  is triggered based on donor characteristics (death cause group,
  malignancy, marginal donor (Y/N), drug abuse, donor weight, age and
  virology) with a Cox proportional hazards model. This model is
  stratified by the donor country and donor blood group.
\item
  offering the graft in order of the match list, while maintaining a
  count of the number of offers made,
\item
  if the maximum number of offers is reached

  \begin{itemize}
  \item
    stop offering the graft further to candidates whose allocation
    profiles exclude rescue donors,
  \item
    giving absolute priority to regional candidates in Germany and
    local candidates in Belgium.
  \end{itemize}
\end{enumerate}

In counting the number of offers made, a center-level rejection is
counted as a single offer, and a candidate-level rejection was counted
only if (a) the graft was compatible with the candidates allocation
profile and (b) the graft was previously rejected by fewer than 5
candidate from the same center.

We point out that this implementation is heavily simplified from actual
rescue allocation in ELAS. Reasons for this are that there exist three
distinct rescue allocation mechanisms which can be triggered under
different circumstances . A second complicating factor is that it is not
clear from graft information alone when competitive rescue allocation is
triggered, nor which centers are eligible for such offers. Moreover, not
all centers are open to accepting grafts in competitive rescue
allocation.

\FloatBarrier
\vspace{.1cm}
\newpage

\section{Simulating post-transplant mortality and re-listing and re-transplantations}\label{app:posttxp}

\subsection{Post-transplant survival models \& relistings}\label{post-transplant-survival-models-relistings}

This section discusses how post-transplant survival and patient
relisting is simulated based on patient and donor characteristics in
ELASS. For post-transplant survival, parametric survival models were
used. The Kaplan-Meier estimator was used to determine the
time-to-relisting relative to simulated post-transplant survival times.
Both models were estimated on transplantations in the Eurotransplant
region between 01-01-2012 and 01-01-2020.

\subsubsection{Choice of outcome: graft failure, patient failure or relisting}\label{choice-of-outcome-graft-failure-patient-failure-or-relisting}

The Eurotransplant registry bi-annually requests data from centra on the
graft failure and patient failure dates for transplanted recipients.
However, centers do not use standardized definitions for graft failure
dates. E.g., some centers define the date of graft failure as the date
on which the patient was re-registered for liver transplantation,
whereas others define the graft failure date as the date on which the
patient was re-transplanted. Additionally, some centers report graft
failure dates for patients even if the patient was not re-listed for
transplantation, or died on the waiting list only years later. This
motivated us to avoid using reported graft failure dates from the
Eurotransplant registry, and instead simulate in ELASS for each
transplanted patient (i) an event time until patient death /
retransplantation (whichever occurred first), and (ii) a
time-to-relisting, relative to the first event time. LSAM handles
post-transplant survival similarly, by assuming that a failure would
result in a patient death unless the candidate is transplanted .

\paragraph{\texorpdfstring{Simulating a time-to-failure \(t\) for transplant recipients}{Simulating a time-to-failure t for transplant recipients}}\label{sec:elassimposttxp}

A Weibull model is used to simulate a time-to-event until patient death or
liver transplantation, separately for HU/ACO and elective candidates.
This Weibull distribution is parametrized with a shape
parameter \(k\) and scale parameter
\(\lambda = \beta^{\T}x\) , where \(x\) are relevant patient and
donor characteristics. The survival function for the Weibull distribution is given by:
\[S(t|x) = \exp\Bigg(-\Big(\frac{t}{\beta^{\T}x}\Big)^k\Bigg).\]

After
obtaining estimates for \(\beta\) and \(k\) based on historical data using Weibull
regression, we can
simulate a time-to-event \(t_i\) for patient-donor pair \(i\) by inverse transform
sampling from this distribution. Specifically, we can draw a random
number \(u \sim \texttt{unif}(0,1)\) and simulate a patient's
time-to-event as
\[t_i = \hat{\mathbf{\beta}}^{\T} \mathbf{x_i} \Big(-\ln(u)^{\frac{1}{k}}\Big).\]

By default, post-transplant survival of elective candidates is simulated based
on a broad set of patient attributes (MELD biomarkers, patient age, country, sex, exception
scores, BMI), donor attributes (year reported, donor age, split, DCD or DBD, death cause,
malignancy, tumor history), and match attributes (weight difference, travel time,
blood group compatibility, match geography). Paucity of data on HU and ACO transplantations
motivated us to adjust for fewer variables in the HU model (weight difference,
donor death cause, donor age, patient age, lab-MELD score, match geography, transplantation
history (Y/N)). For elective candidates, \(t\) is also simulated using country-specific shape
parameters. For HU/ACO candidates, a single shape parameter is used.

\paragraph{\texorpdfstring{Simulating a time-to-relisting \(r\) for transplant recipients}{Simulating a time-to-relisting r for transplant recipients}}\label{sec:elassrelisting}

The simulation of a candidate's time-to-relisting is complicated by the fact that
a patient's time-to-relisting \(r_i\) necessarily has to occur before their
time-to-failure \(t_i\). To simulate re-listing times in the ELAS simulator, we
estimated the probability of re-listing using the Kaplan-Meier estimator, with time elapsed relative to the event time \(t_i\) as the timescale. A
re-listing time can then be simulated by inverse transform sampling from the
Kaplan-Meier curve, i.e.,(i)
sample a random \(u \sim \text{unif}(0,1)\), (ii) choose the first \(s\) such that \(\mathbb{P}[R/T > s] \geq u\), and (iii)
calculating the time-to-relisting as \(s \cdot t\).

\begin{figure}[h]

{\centering \includegraphics[width=1\linewidth]{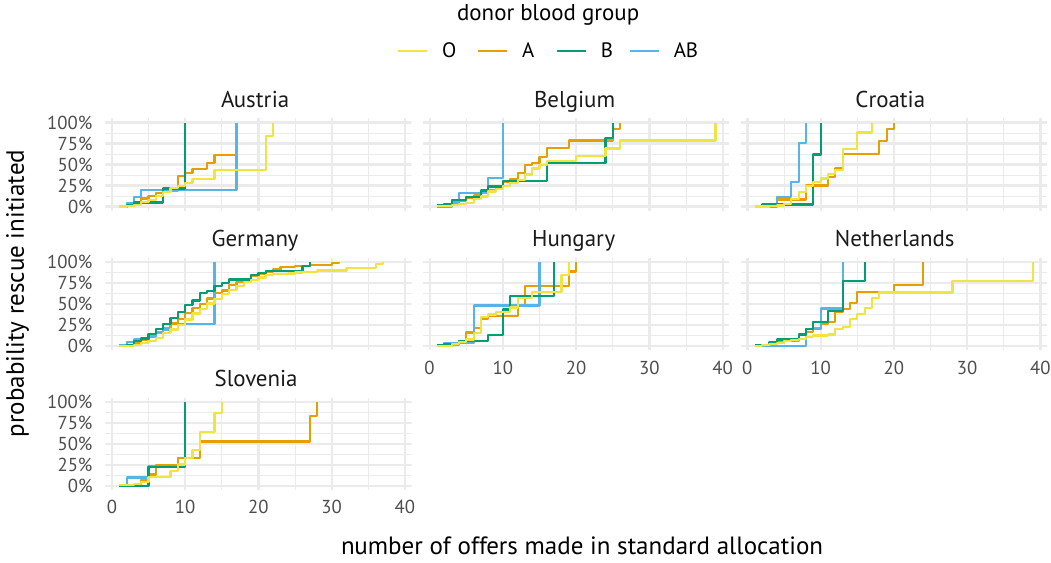} 

}

\caption{Relation between the number of offers made in standard allocation and the probability that non-standard allocation is initiated. Stratum-specific survival curves were generated by predicting the number of offers for an average patient.}\label{fig:sfig5}
\end{figure}

In case no such \(t\) exists, the patient will die without being enlisted for transplantation.
Figure \ref{fig:sfig5} shows estimates of the empirical
distribution of \(R\) relative to \(T\), stratified by discretized \(T\). The figure
shows that the fraction of patients dying after transplantation
without re-listing depends strongly on \(T\). For example, almost 70\% of
candidates who experience an event within 7 days after transplantation are
enlisted for a repeat liver transplantation, whereas only 20\% of candidates
who experience an event more than 5 years after initial transplantation are
enlisted for repeat transplantation.

\paragraph{Constructing synthetic re-registrations}\label{sec:elassynthreg}

If a candidate is re-activated on the waiting list before
the simulation end date, i.e., \(r\) \textless{} \(t\), the post-transplant module adds a \emph{synthetic} repeat
listing to the waiting list. This synthetic listing is
generated by combining the transplant recipient's fixed characteristics with the
dynamic status updates of a candidate who was actually
re-listed for transplantation. This re-listing is chosen such that (a) candidates have
similar time-to-failure \(t\) and similar time-to-relisting \(r\), and (b) candidates
match on pre-determined characteristics.

By default, the post-transplant module finds a matching re-registration
\(k\) by:

\begin{enumerate}
\def\labelenumi{\arabic{enumi}.}
\item
  Considering re-listings where candidates match in terms of whether
  they were re-listed within 14 days\footnote{Such patients can be
    eligible for a HU status under the current allocation rules \citep{ETLiverMan2025}} after transplantation,
  were listed in the same country, are similarly aged (\textless20 years difference),
  and have similar time-to-relistings and time-to-events (\textless1 year difference).
\item
  Selecting the m=5 listings for repeat transplantation with the closest Mahalanobis
  distance between (\(r_i\), \(t_i\)) and (\(r_k\), \(t_k\)).
\item
  Sampling a random re-registration from the \(m\) re-registrations.
\end{enumerate}

A synthetic re-registration is then constructed by combining patient
attributes from patient \(i\) with status updates from patient \(k\). Importantly,
the post-transplant module does not copy over profile status
updates from patient \(k\) to patient \(i\). Awarded exception scores are copied over in case both
patients are from the same country-of-listing because certain (N)SEs are
associated with listing for repeat transplantation (for example, hepatic artery
thrombosis).

\end{document}